\documentclass[prl,aps,epsf,showpacs,twocolumn]{revtex4-1}
\usepackage{amsmath}
\usepackage{amssymb}
\usepackage{amsthm}
\usepackage{amsfonts}%
\usepackage{enumerate}
\usepackage{latexsym}
\usepackage{color}
\usepackage{setspace} 
\usepackage{blindtext}
\usepackage{dsfont}
\usepackage{mathrsfs}
\usepackage[normalem]{ulem}

\usepackage{tabularx}
\newcolumntype{Y}{>{\centering\arraybackslash}X}

\usepackage{stackengine}

\usepackage{bm}
\usepackage{graphicx}
\usepackage{subfigure}

\usepackage{hyperref}
\hypersetup{
pdfnewwindow=true, colorlinks=true,
linkcolor=blue, anchorcolor=blue,
citecolor=blue, filecolor=blue,
menucolor=blue, urlcolor=blue} 

\usepackage{pifont}

\newcommand{\be}[0]{\begin{equation}}
\newcommand{\ee}[0]{\end{equation}}
\newcommand{\ba}[0]{\begin{eqnarray}}
\newcommand{\ea}[0]{\end{eqnarray}}
\newcommand{\mx}[0]{\begin{pmatrix}}
\newcommand{\ex}[0]{\end{pmatrix}}

\newcommand{\Rbf}{\ensuremath{\bm R}}

\newcommand{\kbf}{\ensuremath{\bm k}}

\newcommand{\up}[0]{\uparrow }

\begin{document}

\title{
Dirac zeros in an orbital selective Mott phase:\linebreak
Green's function Berry curvature and flux quantization
}

\author{Lei Chen$^{1}$, Haoyu Hu$^{2,1}$, 
Maia\ G.\ Vergniory$^{2,3}$,
Jennifer Cano$^{4,5}$, 
and Qimiao Si$^{1}$}
\affiliation{$^1$Department of Physics and Astronomy, Rice Center for Quantum Materials, Rice University, Houston, Texas 77005, USA}
\affiliation{$^2$Donostia International  Physics  Center,  P. Manuel  de Lardizabal 4,  20018 Donostia-San Sebastian,  Spain}
\affiliation{$^3$Max Planck Institute for Chemical Physics of Solids, Noethnitzer Str. 40, 01187 Dresden, Germany}
\affiliation{$^4$Department of Physics and Astronomy, Stony Brook University, Stony Brook, NY 11794, USA}
\affiliation{$^5$Center for Computational Quantum Physics, Flatiron Institute, New York, NY 10010, USA}


\begin{abstract}
How electronic topology develops in strongly correlated systems represents a fundamental challenge in the field of quantum materials. 
Recent studies have advanced the characterization and diagnosis of topology in Mott insulators whose underlying electronic structure is topologically nontrivial, through ``Green's function zeros". However, their counterparts in metallic systems have yet to be explored.
Here, we address this problem in an orbital-selective Mott phase (OSMP), which is of extensive interest to a variety of strongly correlated systems with a short-range Coulomb repulsion.
We demonstrate symmetry protected 
crossing of the zeros in an OSMP.
Utilizing the 
concept of Green's function Berry curvature, we show that the zero crossing has a quantized Berry flux. The resulting  notion of Dirac zeros
provides a window into the largely hidden landscape of topological zeros in strongly correlated metallic systems and, moreover, opens up a means to diagnose strongly correlated topology in new materials classes.
\end{abstract}


\maketitle

{\it Introduction.~~} 
Quantum fluctuations as promoted by 
strong correlations drive novel electronic phases of matter
~\cite{Kei17.1,Pas21.1}, and this 
also applies
when 
the underlying electronic structure is topologically nontrivial~\cite{Stormer1999,Xie2021,Zeng-fci2023,Lu-fqahe23.x,Lai2018,Dzsaber2017,Dzs-giant21.1,Chen-Si2022}. One important question is how 
to diagnose electronic topology in
the strongly correlated settings.
In band theory of noninteracting crystalline systems, lattice symmetries constrain the one-electron eigenstates--the Bloch states--and serve as indicators for topology~\cite{Armitage2017,Nagaosa2020, Bradlyn2017,Cano2018,Po2017,Watanabe2017}; 
they uniquely determine the representations~\cite{Cano2018} of the Bloch states and justify the robustness of band degeneracies in the energy dispersion~\cite{cano2021band}.
In interacting systems, a fundamental quantity is the single-particle Green's function~\cite{AGD}, which describes the frequency and wavevector-resolved propagation of an electron in the many-particle environment.
The eigenvectors of the Green's function form 
a representation of the space group~\cite{Hu-Si2021}.
In addition, these eigenvectors allow for the introduction of a Green's function Berry curvature~\cite{Setty_v3}.

Mott insulators develop 
when the Coulomb repulsion is strong (and the band filling is commensurate). They have been known to feature Green's function zeros \cite{Dzyaloshinskii2003,Gurarie2011-2,You-smg2018},
contours in frequency-momentum space at which the Green's function vanishes. The roles of these zeros in the topology of Mott insulators have recently gained considerable interest~\cite{Setty_v1,Sangiovanni_1} in the presence of topological noninteracting 
bands~\cite{Morimoto2016}.
In particular, it was shown that symmetry constraints based on Green's function eigenvectors also act on the zeros~\cite{Setty_v1}.
Moreover, the notion of Green's function Berry flux quantization has been 
illustrated in a Mott insulator derived from long-ranged interactions~\cite{Setty_v3}.
The quantized flux of the zeros
serves as a means to search for topological Mott insulators.

In this work, we address whether and how Green's function zeros can play a role in the topology of correlated metallic systems.
To this end, we study an orbital-selective Mott phase, arising in correlated multi-orbital systems when a subset of the orbitals are (on the verge of being) localized. OSMP and its proximate regimes are
increasingly being
recognized as important to
 a variety of strongly correlated materials platforms \cite{Pas21.1,Kir20.1,Si01.1,Col01.1,Sen04.1,Anisimov2002,Yi17.1,Si-Hussey23,Shan2023,Guerci2022.x,Xie2023.x,Zhao2023.x,Hu-coupled22.3,Hua23.3x}.
Furthermore, we consider a local Coulomb repulsion.
We
show that symmetry protected 
zero crossings occur
and demonstrate the 
 notion of Dirac zeros in terms of
a quantization of 
the Green's function Berry flux.
Given that the orbital-selective Mott physics is prevalent across the strongly correlated metallic systems, 
our work opens up an important new avenue to realize strongly correlated topological matter.

{\it Model and Methods.~~} 
We consider a four-orbital interacting Hamiltonian defined on a cubic lattice, taking the form of $\mathcal{H} = \mathcal{H}_{0} + \mathcal{H}_{int}$. 

The noninteracting part preserves a U(1)-spin rotational symmetry along the $z$-axis;
in accordance, $\mathcal{H}^{\up}_0 = [\mathcal{H}^{\downarrow}_0]^{\dagger}$. The kinetic Hamiltonian for the spin-$\up$ electrons is given by $\mathcal{H}_0^{\up} = \sum_{\kbf} \Psi_{\kbf\up}^{\dagger} H_{0}^{\up}(\kbf) \Psi_{\kbf\up} $,  where $\Psi_{\kbf\up} = [\psi_{Ah\up}, \psi_{Bh\up} \psi_{Al\up}, \psi_{Bl\up}]^{T} $, with subscripts 
$\tau=A/B$ and $o=h/l$ introduced to distinguish the sublattices and heavy/light orbitals, respectively.
\begin{equation}
    H_{0}^{\up}(\kbf) = \mx
    H_{0,hh}^{\up}(\kbf) & H_{0,hl}^{\up}(\kbf) \\
    [H_{0,hl}^{\up}(\kbf)]^{\dagger} & H_{0,ll}^{\up}(\kbf)
    \ex,
\end{equation}
in which $H_{0,hh/ll}^{\up}(\kbf)= 2t_{h}/t_{l} (\cos k_x + \cos k_y + \cos k_z)\tau_z + 2t_{h/l}^{soc} (\sin k_x \tau_x + \sin k_y \tau_y)$ and $H_{0, hl}^{\up}(\kbf) =2t_{hl} (\cos k_x + \cos k_y + \cos k_z) $.  $\tau_{x/y/z}$ are the Pauli matrices acting on the sublattice space. Furthermore, $t_h$ ($t_l$) and $t_h^{soc}$ ($t_l^{soc}$) denote the nearest neighbor direct hopping and spin-orbit coupling between the heavy (light) orbitals, respectively, while $t_{hl}$ represents the nearest neighbor hopping between the heavy and light orbitals.  
We 
consider 
$|t_{h}|$ to be considerably smaller than $|t_l|$, 
such that for the same interaction strength,
the heavy orbitals are much more correlated than the light orbitals. 
Without a loss of generality, we adopt the following parameters:
$t_h =1$, $t_h^{soc}=0.6$, $t_l =4$, $t_l^{soc}=2.4$ and $t_{hl}=1.5$. 

We will consider the effect of an onsite Coulomb repulsion. Our focus will be on a range of interactions that are small compared to the light bandwidth and, accordingly, have a relatively weak effect on the light band. Thus, for simplicity, we will study only the effect of
an intraorbital  Hubbard interaction that  acts on the heavy orbitals: 
\begin{equation}
    \mathcal{H}_{int} = \sum_{\tau} \frac{U}{2}\left( n_{\tau h\up} + n_{\tau h\downarrow} -1\right)^2 \,,
\end{equation}
where the index $\tau$ enumerates the two subalttices of the heavy orbitals. 
The model contains a $C_{4z}$ rotational symmetry, which protects the gapless Dirac points in the weak coupling region.

Dispersive zeros may arise only if the electron self-energy is ${\bf k}$-dependent. To capture such effects, we develop a cluster version of the U(1)-slave-spin (SS) method \cite{Yu_ss}.
In 
the U(1)-SS approach, 
an electronic operator is expressed
as $\psi_{i\tau o\sigma}^{\dagger} = S^+_{i\tau o\sigma}f_{i\tau o\sigma}^{\dagger}$, where the SS operator $S^+$ is introduced to carry the charge degree of freedom and the auxiliary fermion $f^{\dagger}$ 
the spin degree of freedom in the form of a ``spinon" operator. The enlarged Hilbert space spanned by the 
SS and auxiliary fermion is limited to the physical ones when we impose the constraint $S_{i\tau o\sigma}^{z} +1/2 = f^{\dagger}_{i\tau o\sigma}f_{i\tau o\sigma} $. We treat the Hamiltonian in the SS 
representation 
at the saddle-point 
level,
which leads to two decoupled effective Hamiltonians, each containing only
the SS or auxiliary fermions,
respectively
[See the supplementary materials (SM)]. Furthermore, we exactly diagonalize the SS Hamiltonian which is constrained on a finite cluster
that is embedded in an 
effective medium; here, we consider a cluster size of two unit cells.

The Mott transition of an orbital is signaled by the vanishing of its quasiparticle weight, which is 
determined from $Z_{i\tau o\sigma} = |\langle S_{i\tau o\sigma}^{+}\rangle|^2$. 
The spatial fluctuations are captured by the bond operators $\chi^{f}_{i\tau o\sigma, j\tau' o'\sigma'} =\langle S^{+}_{i\tau o\sigma}S^{-}_{j\tau'o'\sigma'}\rangle$. Further details of the method and its context
are given in the SM.

\begin{figure}[t!]
    \centering
    \includegraphics[width=1.05\linewidth]{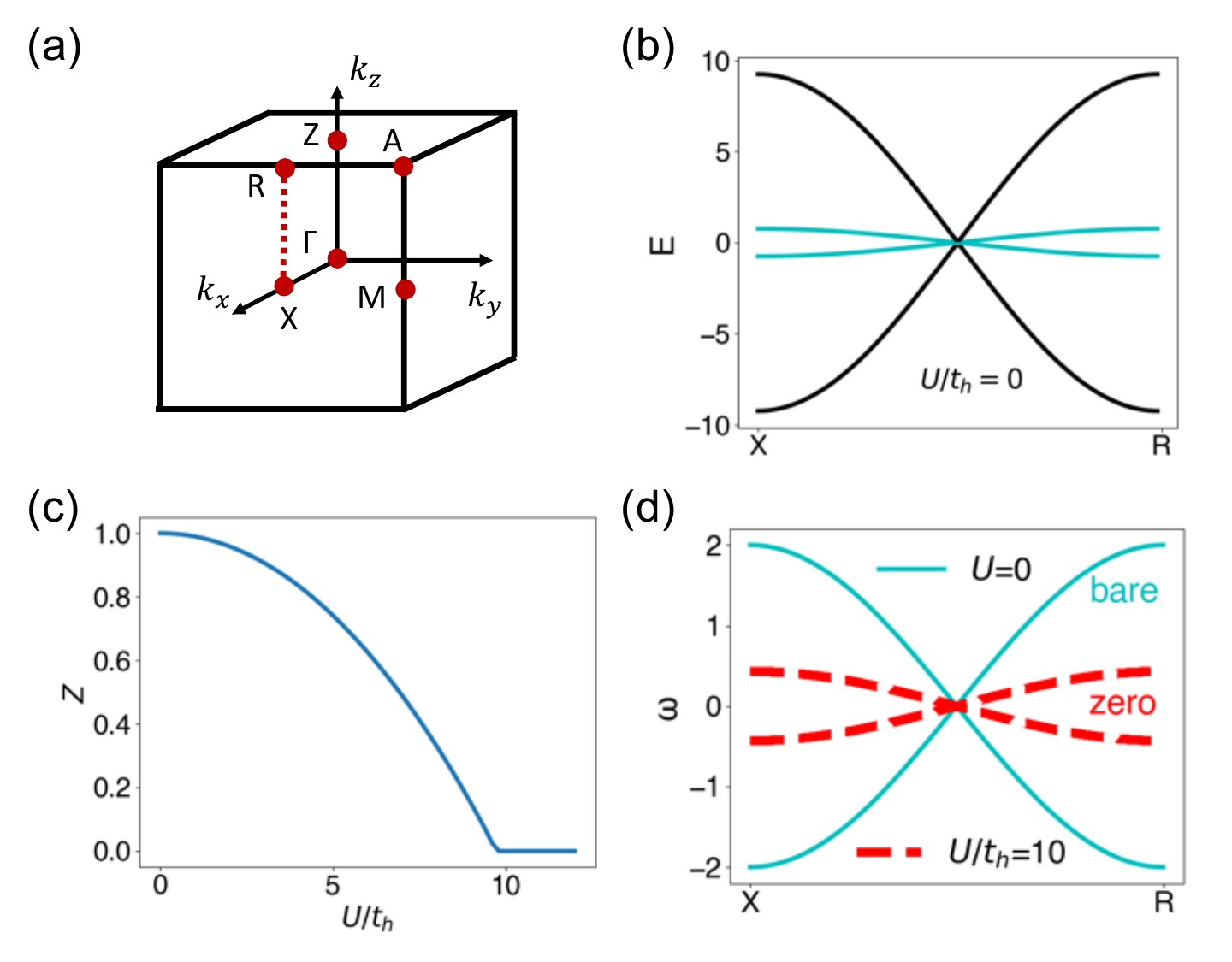}
    \caption{ 
   (a) Brillouin zone of a cubic lattice, with high symmetry points marked. (b) The non-interacting band structure of the narrow (cyan) and wide (black) bands along a high symmetry line ($X-R$),  showing a Dirac node protected by symmetry.
 (c) Quasiparticle weight of the heavy electrons as a function of the onsite interaction $U$. (d) Red dashed lines: Green's function zeros in the OSMP phase, for $U/t_h=10$,
    dispersing along $X-R$. Cyan solid lines: Dispersion of the bare narrower-band electrons,
    at $U=0$.
    }
    \label{fig:qp}
\end{figure}

{\it Orbital selective Mott phase.~~}
In the noninteracting limit, the Hamiltonian describes 
two species of semimetals, each associated with one kind of orbital and each having
four gapless (Dirac) points sitting at the momenta $(\pi, 0, \pm \frac{\pi}{2})$ and $(0,\pi, \pm \frac{\pi}{2})$. The system contains both the time-reversal ($\mathcal{T}$) and inversion ($\mathcal{I}$) symmetries. 
The discrete rotation 
$C_{4z}$ symmetry 
protects the Dirac crossings along the high symmetry lines paralleled to the $z$-axis. The noninteracting band structure along the high symmetry line $X-R$ is displayed in Fig.~\ref{fig:qp}(b), showing a symmetry protected Dirac node.

We are now in position to discuss the correlation effect. As depicted in Fig.~\ref{fig:qp}(c), the quasiparticle weight of the heavy orbitals monotonically decreases as we increase the strength of the intra-orbital on-site Hubbard interaction.
It eventually diminishes to zero at a critical value $U_c/t_{h}=9.5$. Notice that, the effective interorbital hybridization between the heavy and light electrons equals to $\sqrt{Z_{h}}t_{hl}$, which 
also vanishes
in the OSMP. In other words, the heavy electrons are
decoupled
from the light ones in the OSMP, leading to the notion of correlation-driven dehybridization, which is analogous to  
the Kondo-destroyed fixed point 
of heavy fermion systems
\cite{Si01.1,Col01.1,Sen04.1}.
Such a dehybridization fixed point is realized from the competition between the hybridization and spatial correlations, which protects the stability of the OSMP \cite{Yu-osmp17,Komijani-osmp17}.

{\it Dirac nodes of Green's function zeros.~}
We now turn to 
the single particle Green's function.
In the SS approach, it is represented by the SS and auxiliary fermions as $G_{\sigma}^{\tau o\tau'o'} (i-j, \tau)=-\langle T_{\tau} S_{i\tau o\sigma}^{-}(\tau) S_{j\tau' o'\sigma}^{+}(0) f_{i\tau o\sigma}(\tau)f_{j\tau' o'\sigma}^{\dagger} (0) \rangle$. At the saddle-point level, 
the Green's function is decomposed into coherent and incoherent parts as follows:
\begin{equation}\label{eq:g_ss}
\begin{aligned}
     G^{\tau o\tau'o'}(\kbf, i\omega_n) &= \sqrt{Z_{\tau o}Z_{\tau' o'}} G^{\tau o\tau'o'}_{f}(\kbf, i\omega_n) \\
     &+ G_{incoh}^{\tau o\tau'o'} (\kbf, i\omega_n) \, .
\end{aligned}
\end{equation}
 In the coherent part, $G_{f}^{\tau o\tau'o'}$ represents the Green's function of the auxiliary fermion. In our case, 
the quasiparticle weights ($Z_{\tau l}$) for the light orbitals are equal to $1$ across the whole phase diagram. In the OSMP phase, the heavy electrons completely lose their coherence
 and, thus, the coherent part of the Green's function is entirely comprised of the light orbitals. On the other hand, the incoherent part of the Green's function is uniquely contributed by the heavy orbitals.
 The momentum-resolved incoherent Green's function is obtained by a Fourier transform of 
 the real space counterpart. 
This leads to
 \begin{equation}
 \begin{aligned}
      G_{incoh}^{\tau o\tau'o'} (\kbf, i\omega_n) &= G^{\tau o\tau'o'}_{incoh}(\bm{0}, i\omega_n) \\
      &+ \sum_{\Rbf \neq \bm{0}
      } e^{i\kbf \cdot \Rbf}G_{incoh}^{\tau o\tau'o'} (\Rbf, i\omega_n) \, ,
 \end{aligned}
 \end{equation}
 where $\Rbf$ enumerates through the separations between the 
 unit cells.
 The specific expressions of $G^{\tau o\tau'o'}_{incoh}(\bm{0}, i\omega_n)$ and $G_{incoh}^{\tau o\tau'o'} (\Rbf, i\omega_n)$ are 
 shown in the SM. 

 We focus on the OSMP, which is illustrated with the choice of 
 the interaction strength $U/t_h=10$.
 We calculate the determinant of the retarded Green's function  $|\det G (\kbf, \omega +i0^{+})|$ for the heavy orbitals in order to look for the zeros of the interacting single-particle Green's function. 
 
\begin{figure}[t!]
    \centering
\includegraphics[width=1.05\linewidth]{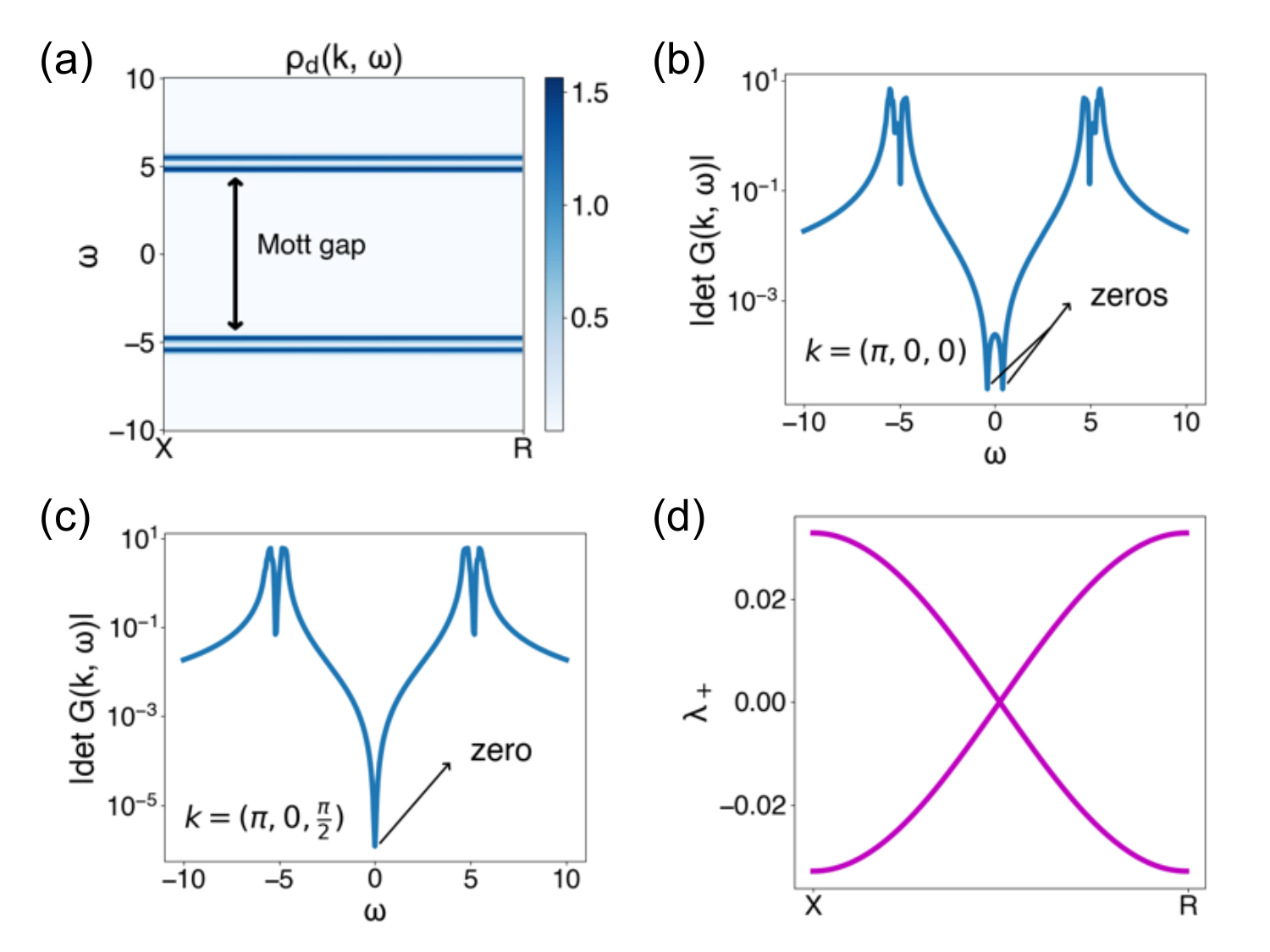}
    \caption{
  (a) The energy distribution of the total spectral function along
    $X-R$, at $U/t_h=10$. The Mott gap is marked.
    (b) and (c),
    The variance of $|\det G(\kbf, \omega)|$ for the heavy orbitals versus $\omega$, demonstrating the Green's function zeros,
    at $\kbf=(\pi, 0, 0)$ and $\kbf=(\pi, 0, \pi/2)$, respectively, 
    (d) Eigenvalues of $G_{+}(\kbf, \omega)$ at fixed frequency $\omega=0$, along the high symmetry line $X-R$.
    }
    \label{fig:detG}
\end{figure}

To further elaborate on the electronic properties in the OSMP, we show the spectrum function for the heavy orbitals in Fig.~\ref{fig:detG}(a), which is calculated from $\rho^h(\kbf, \omega) = -\frac{1}{\pi}\sum_{\tau \sigma}G^{\tau h\tau h}_{\sigma}(\kbf, \omega+i0^{+})$.
The lower and upper Hubbard bands are separated by a Mott gap as marked by the arrow shown in Fig.~\ref{fig:detG}(a).
We then turn to 
discussing the frequency resolved determinant of the 
Green's function. Its frequency dependence at  $\kbf=(\pi, 0, 0)$ and $\kbf=(\pi,0,\frac{\pi}{2})$ 
is respectively displayed in Figs.~\ref{fig:detG}(b,c) 
on a logarithmic scale. 
The lower and upper Hubbard bands
correspond to
the broad shoulder-like peaks at high frequencies.
Whereas the Green's function zeros, marked by the black arrows, reside within the Mott gap. Different from the case of momentum $X$, which is shown in Fig.~\ref{fig:detG}(b), only one dip is observed at momentum $\kbf=(\pi,0,\pi/2)$ 
[Fig.~\ref{fig:detG}(c)], 
showing that the Green's function zeros are merged at this momentum.

The collection of zeros at the different wavevectors 
is depicted in Fig.~\ref{fig:qp}(d). The red dashed lines show the dispersive 
zeros along the high symmetry line $X-R$. Furthermore, two branches of Green's function zeros merge exactly at 
momentum for the Dirac node of the bare heavy electrons in the noninteracting limit, which are denoted by the 
cyan solid lines in 
Fig.~\ref{fig:qp}(d). (Here,
``bare'' means both at $U=0$ and without a hybridization between the heavy and light electrons.) This result demonstrates the presence and robustness of the ``gapless'' node coming from the Green's function zeros in the strongly correlated limit, and motivates us to use the eigenvectors of the Green's function zeros (instead of Green's function poles) to diagnose the topology in such strongly correlated
systems. 

{\it Quantization of Green's function Berry flux.~~}
In the noninteracting limit, 
the winding of the Bloch functions in the momentum space is used to calculate the Berry flux quantization.  
The question is how to define an appropriate Berry phase for  the strongly
correlated case. 
Here, we use 
a frequency-dependent Berry curvature from the Green's function  eigenvectors, which has recently been introduced~\cite{Setty_v3}.

We define a Hermitian combination:
\begin{equation}\label{eq:Gplus}
    G_{+}(\kbf, \omega) = G^{R}(\kbf, \omega) + G^{A}(\kbf, \omega) \, ,
\end{equation}
where $G^{R,A}$ are the retarded and advanced Green's functions.
We focus on this combination (as opposed to, say, $G_{+}^{-1}$) because it is regular across the zeros in the frequency-momentum space.
The $\alpha$-th eigenvalues and eigenvectors of the new Green's function are obtained by the eigenvalue equation,
\begin{equation}
    G_{+}(\kbf, \omega) | \phi_{\alpha} (\kbf, \omega)\rangle = \lambda_{+}^{\alpha} | \phi_{\alpha} (\kbf, \omega)\rangle \, .
\end{equation}
Because of the Hermiticity, the eigenvalues $\lambda_{+}$ are always real. The eigenvalues of $G_{+}$ at $\omega=0$ is depicted in Fig.~\ref{fig:detG}(d). Similar to the noninteracting bands and Green's function zeros, along the same high symmetry line, the ``band''s  show a nodal crossing at the middle point of $X-R$. 

\begin{figure}[t!]
    \centering
    \includegraphics[width=\linewidth]{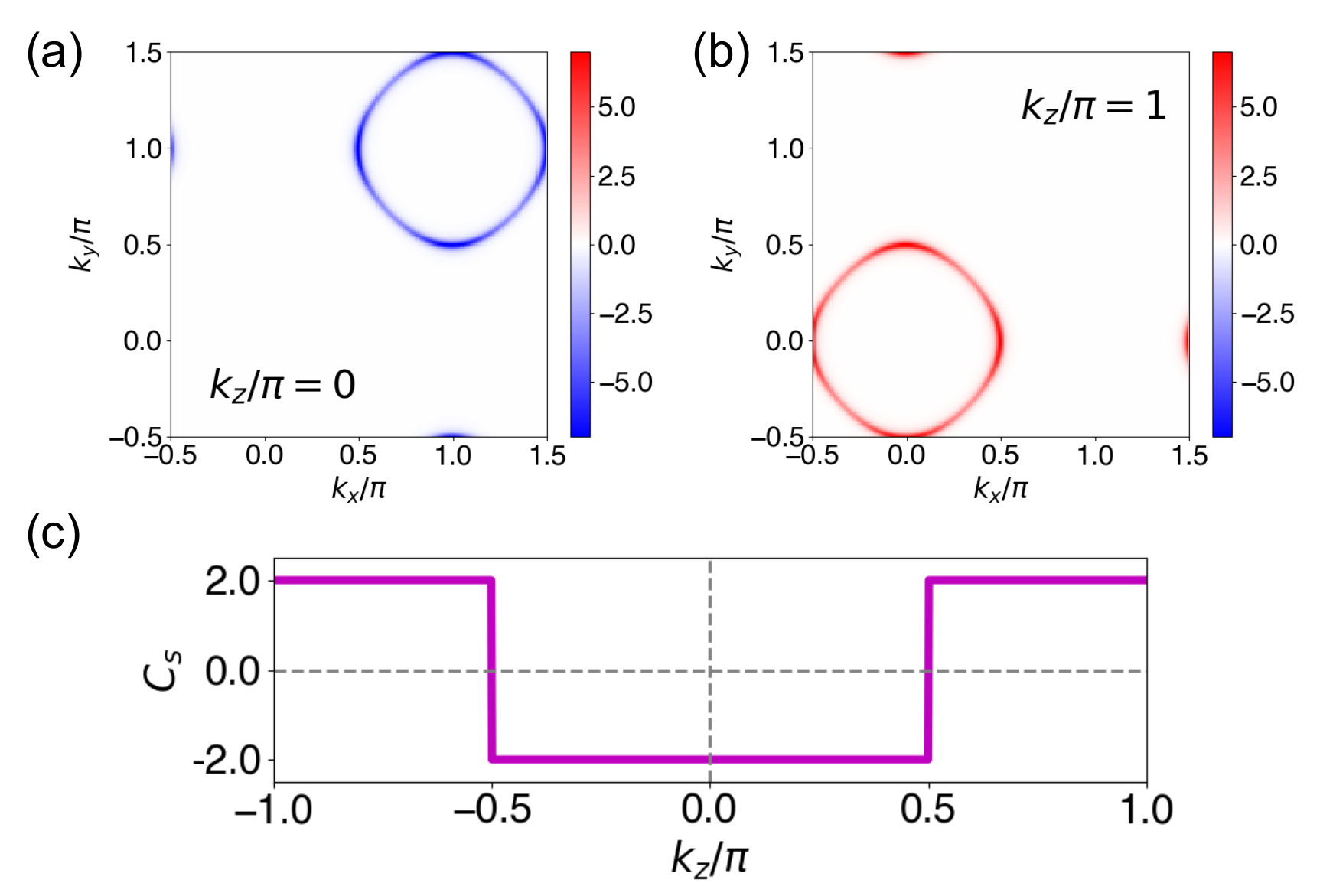}
    \caption{
  (a) and (b),
    The Green's function Berry curvature, $B_{z}^{\up}(\kbf,\omega) $, for frequency $\omega=0$, in the $k_z=0$ and $k_z=\pi$ planes, respectively. (c) The spin Chern number obtained from the Green's function Berry curvature as a variance of $k_z$ at 
    the fixed frequency $\omega=0$, demonstrating the quantization of the corresponding Berry flux around the Dirac node of zeros.
    }
    \label{fig:berry_curv}
\end{figure}

Furthermore, 
from the ``energy'' (eigenvalues $\lambda_{+}$) of the ``bands'', 
we are able to define the Berry curvature associated with the eigenvectors at momentum $\kbf$ and freqeuncy $\omega$. It takes the form of
\begin{equation}
    B_{m} (\kbf, \omega) = \sum_{\alpha, ij} \epsilon_{ijm} \left[ \langle \partial_i \phi_{\alpha} (\kbf, \omega) | \partial_{j} \phi_{\alpha} (\kbf, \omega) \rangle \right]
    \, , 
\end{equation}
where $\alpha$ sums over the ``bands'' with eigenvalue $\lambda_{+}^{\alpha}$ smaller than $0$, and $\epsilon_{ijm}$ is the Levi-Civita symbol. For a given $k_z$ and $\omega$, a frequency dependent spin Chern number is then calculated by
\begin{equation}
    C_s(\omega, k_z) = \frac{i}{2\pi} \sum_{\sigma} (-1)^{\sigma}\int d \kbf_{\perp} B_{z}^{\sigma} (\kbf_{\perp}, k_z, \omega) \, .
\end{equation}
with $\kbf_{\perp} = (k_x, k_y)$, and $\sigma=0(1)$ for spin-$\up$($\downarrow$).

We focus on the case
$\omega=0$. 
The Berry curvatures (spin-$\up$) at $k_z=0$ and $k_z=\pi$ planes are displayed in Figs.~\ref{fig:berry_curv}(a,b), which are seen to have different signs. 
Their distribution in the $k_x-k_y$ plane captures the 
dispersion of the Green's function ``bands".
We find them to be quantized in both planes after integrating the Berry curvature over the whole Brillouin zone plane. We further scan the Chern numbers $C_s$ as a function of $k_z$, which are shown in 
Fig.~\ref{fig:berry_curv}(c). 
There is a sudden jump by a value of $4$, when the cutting plane crosses the nodal points at $k_z=\pm \pi/2$.
There are two nodes at each of the 
$k_z=\pi/2$ and 
$k_z=-\pi/2$ planes (as can be inferred from the SM, Fig.~\ref{fig:nodes}).
Accordingly, there is a jump of $2$ across each crossing.

The quantization of the Berry flux shows an enclosed monopole charge at a touching point of the Green's function bands.
Given that the zero crossing has a four-fold degeneracy and a quantized Green's function monopole charge, we will refer to it as a Dirac zero.

{\it Discussion.~~}
Several remarks are 
in order.
Firstly, by using the Green's function Berry flux, we have defined the notion of Dirac zeros. 
 We expect that this can be readily extended to other settings, such as Weyl zeros in correlated systems with broken inversion or time-reversal symmetry. 
 Secondly, our work provides a systematic approach to diagnose the electronic topology 
in the correlation-driven orbital selective Mott phase.
In addition to heavy fermion metals,
orbital-selective Mott correlations have been extensively discussed in Fe-based superconductors
\cite{Yi17.1,Si-Hussey23}  
and are also being recognized 
in both the moir\'{e} structures~\cite{Shan2023,Guerci2022.x,Xie2023.x,Zhao2023.x} and 
geometry-induced flat band compounds 
\cite{Hu-coupled22.3,Hua23.3x}.
Many of these systems contain 
 topological electron bands. As such, the notion of Dirac/Weyl zeros serves as a diagnostic tool to search and obtain new types of strongly correlated topological phases and materials. 
 Finally, it has been 
 increasingly recognized that the Green’s function zeros contribute to measurable physical quantities, such as the Luttinger volume~\cite{Yunoki2017,Setty_v2},
 Hall effect~\cite{Setty_v2,Blason2023,Gavensky2023,Phillips2023} and
 even quantum oscillations~\cite{Fabrizio2022,Setty_v3} in a consistent way~\cite{Setty_v2}. 
Consequently, the topological feature defined from Green’s function zeros is expected to influence physical properties in a nontrivial manner. This exciting prospect warrants 
further exploration.

 {\it Conclusion.~~} 
 In a multiorbital model with topological bandstructure, we have demonstrated a 
 symmetry protected 
crossing of 
Green's function zeros
in its orbital-selective Mott phase (OSMP).
Through the concept of Green's function Berry curvature, we have found that the four-fold zero crossing carries a quantized monopole charge.
Our results show that the Dirac zeros provide a means to search for and realize new type of topological phases in strongly correlated metallic materials. More generally, our work illustrates the power of strong correlations to enable 
novel topological phases of matter 
in a broad range of quantum materials.

 \textit{Note added:} After the completion of this work,
 recent works addressing 
 different models with a focus on 
 edge zeros in strongly interacting topological insulators became available \cite{Sangiovanni_2,Bollmann}.

{\it Acknowledgement.~~}
We thank Yuan Fang, Elio K\"onig, Gabriela Oprea, Silke Paschen, Chandan Setty, Shouvik Sur, and Fang Xie,
for useful discussions.
Work at Rice has primarily been supported by the Air Force Office of Scientific Research under Grant No.
FA9550-21-1-0356 (model construction,
L.C., H.H. and Q.S.), 
by the National Science Foundation
under Grant No. DMR-2220603 
and the Robert A. Welch Foundation Grant No. C-1411 (model calculation, L.C. and H.H.), and by the Vannevar Bush Faculty Fellowship ONR-VB N00014-23-1-2870 (conceptualization, Q.S.). The
majority of the computational calculations have been performed on the Shared University Grid
at Rice funded by NSF under Grant EIA-0216467, a partnership between Rice University, Sun
Microsystems, and Sigma Solutions, Inc., the Big-Data Private-Cloud Research Cyberinfrastructure
MRI-award funded by NSF under Grant No. CNS-1338099, and the Extreme Science and
Engineering Discovery Environment (XSEDE) by NSF under Grant No. DMR170109. 
H.H. acknowledges
the support of the European Research Council (ERC) under the European Union's 
Horizon 2020 research and innovation program (Grant Agreement No.\ 101020833).
M.G.V. acknowledges support to the  Spanish Ministerio de Ciencia e Innovacion (grant PID2022-142008NB-I00), partial support from European Research Council (ERC) grant agreement no. 101020833 and the European Union NextGenerationEU/PRTR-C17.I1, 
by the IKUR Strategy under the collaboration agreement between Ikerbasque Foundation and DIPC on behalf of the Department of Education of the Basque Government.
as well as by the funding from the Deutsche Forschungsgemeinschaft (DFG, German Research Foundation)
through the project FOR 5249 (QUAST).
J.C. acknowledges the support of
the National Science Foundation under Grant No. DMR-1942447, support from the Alfred P.
Sloan Foundation through a Sloan Research Fellowship and the support of the Flatiron Institute,
a division of the Simons Foundation. 
All authors acknowledge 
the hospitality of the Kavli Institute for Theoretical Physics, UCSB,
supported in part
by the National Science Foundation under Grant No. NSF PHY-1748958,
 during the program ``A Quantum Universe in
a Crystal: Symmetry and Topology across the Correlation Spectrum." 
J.C. and Q.S. also 
acknowledge the hospitality of the Aspen Center for Physics, which is supported by the National Science Foundation under Grant No. PHY-2210452.


\bibliography{ss_zero.bib}

\begin{thebibliography}{56}%
\makeatletter
\providecommand \@ifxundefined [1]{%
 \@ifx{#1\undefined}
}%
\providecommand \@ifnum [1]{%
 \ifnum #1\expandafter \@firstoftwo
 \else \expandafter \@secondoftwo
 \fi
}%
\providecommand \@ifx [1]{%
 \ifx #1\expandafter \@firstoftwo
 \else \expandafter \@secondoftwo
 \fi
}%
\providecommand \natexlab [1]{#1}%
\providecommand \enquote  [1]{``#1''}%
\providecommand \bibnamefont  [1]{#1}%
\providecommand \bibfnamefont [1]{#1}%
\providecommand \citenamefont [1]{#1}%
\providecommand \href@noop [0]{\@secondoftwo}%
\providecommand \href [0]{\begingroup \@sanitize@url \@href}%
\providecommand \@href[1]{\@@startlink{#1}\@@href}%
\providecommand \@@href[1]{\endgroup#1\@@endlink}%
\providecommand \@sanitize@url [0]{\catcode `\\12\catcode `\$12\catcode
  `\&12\catcode `\#12\catcode `\^12\catcode `\_12\catcode `\%12\relax}%
\providecommand \@@startlink[1]{}%
\providecommand \@@endlink[0]{}%
\providecommand \url  [0]{\begingroup\@sanitize@url \@url }%
\providecommand \@url [1]{\endgroup\@href {#1}{\urlprefix }}%
\providecommand \urlprefix  [0]{URL }%
\providecommand \Eprint [0]{\href }%
\providecommand \doibase [0]{http://dx.doi.org/}%
\providecommand \selectlanguage [0]{\@gobble}%
\providecommand \bibinfo  [0]{\@secondoftwo}%
\providecommand \bibfield  [0]{\@secondoftwo}%
\providecommand \translation [1]{[#1]}%
\providecommand \BibitemOpen [0]{}%
\providecommand \bibitemStop [0]{}%
\providecommand \bibitemNoStop [0]{.\EOS\space}%
\providecommand \EOS [0]{\spacefactor3000\relax}%
\providecommand \BibitemShut  [1]{\csname bibitem#1\endcsname}%
\let\auto@bib@innerbib\@empty
\bibitem [{\citenamefont {Keimer}\ and\ \citenamefont {Moore}(2017)}]{Kei17.1}%
  \BibitemOpen
  \bibfield  {author} {\bibinfo {author} {\bibfnamefont {B.}~\bibnamefont
  {Keimer}}\ and\ \bibinfo {author} {\bibfnamefont {J.~E.}\ \bibnamefont
  {Moore}},\ }\href {\doibase
  https://urldefense.proofpoint.com/v2/url?u=http-3A__dx.doi.org_10.1038_nphys4302&d=DwIFAg&c=sJ6xIWYx-zLMB3EPkvcnVg&r=46ZiAColLCC0f_qMjSjIHU-3dYH0tEwJaD76OrlPzkA&m=Zh9nyCUbX1MYa3vNS4Cay_8qMSzjQ3Gv8Wew-zvImpc&s=zbm5aVFQgffvqHZkAWqEGdfkdgU6edtrZCJuxezBNr4&e=}
  {\bibfield  {journal} {\bibinfo  {journal} {{Nat.\ Phys.}}\ }\textbf
  {\bibinfo {volume} {13}},\ \bibinfo {pages} {1045} (\bibinfo {year}
  {2017})}\BibitemShut {NoStop}%
\bibitem [{\citenamefont {Paschen}\ and\ \citenamefont {Si}(2021)}]{Pas21.1}%
  \BibitemOpen
  \bibfield  {author} {\bibinfo {author} {\bibfnamefont {S.}~\bibnamefont
  {Paschen}}\ and\ \bibinfo {author} {\bibfnamefont {Q.}~\bibnamefont {Si}},\
  }\href {\doibase 10.1038/s42254-020-00262-6} {\bibfield  {journal} {\bibinfo
  {journal} {Nat.\ Rev.\ Phys.}\ }\textbf {\bibinfo {volume} {3}},\ \bibinfo
  {pages} {9} (\bibinfo {year} {2021})}\BibitemShut {NoStop}%
\bibitem [{\citenamefont {Stormer}\ \emph {et~al.}(1999)\citenamefont
  {Stormer}, \citenamefont {Tsui},\ and\ \citenamefont
  {Gossard}}]{Stormer1999}%
  \BibitemOpen
  \bibfield  {author} {\bibinfo {author} {\bibfnamefont {H.~L.}\ \bibnamefont
  {Stormer}}, \bibinfo {author} {\bibfnamefont {D.~C.}\ \bibnamefont {Tsui}}, \
  and\ \bibinfo {author} {\bibfnamefont {A.~C.}\ \bibnamefont {Gossard}},\
  }\href {\doibase 10.1103/RevModPhys.71.S298} {\bibfield  {journal} {\bibinfo
  {journal} {Rev. Mod. Phys.}\ }\textbf {\bibinfo {volume} {71}},\ \bibinfo
  {pages} {S298} (\bibinfo {year} {1999})}\BibitemShut {NoStop}%
\bibitem [{\citenamefont {Xie}\ \emph {et~al.}(2021)\citenamefont {Xie},
  \citenamefont {Pierce}, \citenamefont {Park}, \citenamefont {Parker},
  \citenamefont {Khalaf}, \citenamefont {Ledwith}, \citenamefont {Cao},
  \citenamefont {Lee}, \citenamefont {Chen}, \citenamefont {Forrester},
  \citenamefont {Watanabe}, \citenamefont {Taniguchi}, \citenamefont
  {Vishwanath}, \citenamefont {Jarillo-Herrero},\ and\ \citenamefont
  {Yacoby}}]{Xie2021}%
  \BibitemOpen
  \bibfield  {author} {\bibinfo {author} {\bibfnamefont {Y.}~\bibnamefont
  {Xie}}, \bibinfo {author} {\bibfnamefont {A.~T.}\ \bibnamefont {Pierce}},
  \bibinfo {author} {\bibfnamefont {J.~M.}\ \bibnamefont {Park}}, \bibinfo
  {author} {\bibfnamefont {D.~E.}\ \bibnamefont {Parker}}, \bibinfo {author}
  {\bibfnamefont {E.}~\bibnamefont {Khalaf}}, \bibinfo {author} {\bibfnamefont
  {P.}~\bibnamefont {Ledwith}}, \bibinfo {author} {\bibfnamefont
  {Y.}~\bibnamefont {Cao}}, \bibinfo {author} {\bibfnamefont {S.~H.}\
  \bibnamefont {Lee}}, \bibinfo {author} {\bibfnamefont {S.}~\bibnamefont
  {Chen}}, \bibinfo {author} {\bibfnamefont {P.~R.}\ \bibnamefont {Forrester}},
  \bibinfo {author} {\bibfnamefont {K.}~\bibnamefont {Watanabe}}, \bibinfo
  {author} {\bibfnamefont {T.}~\bibnamefont {Taniguchi}}, \bibinfo {author}
  {\bibfnamefont {A.}~\bibnamefont {Vishwanath}}, \bibinfo {author}
  {\bibfnamefont {P.}~\bibnamefont {Jarillo-Herrero}}, \ and\ \bibinfo {author}
  {\bibfnamefont {A.}~\bibnamefont {Yacoby}},\ }\href {\doibase
  10.1038/s41586-021-04002-3} {\bibfield  {journal} {\bibinfo  {journal}
  {{Nature}}\ }\textbf {\bibinfo {volume} {600}},\ \bibinfo {pages} {439}
  (\bibinfo {year} {2021})}\BibitemShut {NoStop}%
\bibitem [{\citenamefont {Zeng}\ \emph {et~al.}(2023)\citenamefont {Zeng},
  \citenamefont {Xia}, \citenamefont {Kang}, \citenamefont {Zhu}, \citenamefont
  {Kn\"uppel}, \citenamefont {Vaswani}, \citenamefont {Watanabe}, \citenamefont
  {Taniguchi}, \citenamefont {Mak},\ and\ \citenamefont {Shan}}]{Zeng-fci2023}%
  \BibitemOpen
  \bibfield  {author} {\bibinfo {author} {\bibfnamefont {Y.}~\bibnamefont
  {Zeng}}, \bibinfo {author} {\bibfnamefont {Z.}~\bibnamefont {Xia}}, \bibinfo
  {author} {\bibfnamefont {K.}~\bibnamefont {Kang}}, \bibinfo {author}
  {\bibfnamefont {J.}~\bibnamefont {Zhu}}, \bibinfo {author} {\bibfnamefont
  {P.}~\bibnamefont {Kn\"uppel}}, \bibinfo {author} {\bibfnamefont
  {C.}~\bibnamefont {Vaswani}}, \bibinfo {author} {\bibfnamefont
  {K.}~\bibnamefont {Watanabe}}, \bibinfo {author} {\bibfnamefont
  {T.}~\bibnamefont {Taniguchi}}, \bibinfo {author} {\bibfnamefont {K.~F.}\
  \bibnamefont {Mak}}, \ and\ \bibinfo {author} {\bibfnamefont
  {J.}~\bibnamefont {Shan}},\ }\href {\doibase 10.1038/s41586-023-06452-3}
  {\bibfield  {journal} {\bibinfo  {journal} {{Nature}}\ }\textbf {\bibinfo
  {volume} {622}},\ \bibinfo {pages} {69} (\bibinfo {year} {2023})}\BibitemShut
  {NoStop}%
\bibitem [{\citenamefont {Lu}\ \emph {et~al.}(2023)\citenamefont {Lu},
  \citenamefont {Han}, \citenamefont {Yao}, \citenamefont {Reddy},
  \citenamefont {Yang}, \citenamefont {Seo}, \citenamefont {Watanabe},
  \citenamefont {Taniguchi}, \citenamefont {Fu},\ and\ \citenamefont
  {Ju}}]{Lu-fqahe23.x}%
  \BibitemOpen
  \bibfield  {author} {\bibinfo {author} {\bibfnamefont {Z.}~\bibnamefont
  {Lu}}, \bibinfo {author} {\bibfnamefont {T.}~\bibnamefont {Han}}, \bibinfo
  {author} {\bibfnamefont {Y.}~\bibnamefont {Yao}}, \bibinfo {author}
  {\bibfnamefont {A.~P.}\ \bibnamefont {Reddy}}, \bibinfo {author}
  {\bibfnamefont {J.}~\bibnamefont {Yang}}, \bibinfo {author} {\bibfnamefont
  {J.}~\bibnamefont {Seo}}, \bibinfo {author} {\bibfnamefont {K.}~\bibnamefont
  {Watanabe}}, \bibinfo {author} {\bibfnamefont {T.}~\bibnamefont {Taniguchi}},
  \bibinfo {author} {\bibfnamefont {L.}~\bibnamefont {Fu}}, \ and\ \bibinfo
  {author} {\bibfnamefont {L.}~\bibnamefont {Ju}},\ }\href@noop {} {\bibfield
  {journal} {\bibinfo  {journal} {arXiv preprint arXiv:2309.17436}\ } (\bibinfo
  {year} {2023})}\BibitemShut {NoStop}%
\bibitem [{\citenamefont {Lai}\ \emph {et~al.}(2018)\citenamefont {Lai},
  \citenamefont {Grefe}, \citenamefont {Paschen},\ and\ \citenamefont
  {Si}}]{Lai2018}%
  \BibitemOpen
  \bibfield  {author} {\bibinfo {author} {\bibfnamefont {H.-H.}\ \bibnamefont
  {Lai}}, \bibinfo {author} {\bibfnamefont {S.~E.}\ \bibnamefont {Grefe}},
  \bibinfo {author} {\bibfnamefont {S.}~\bibnamefont {Paschen}}, \ and\
  \bibinfo {author} {\bibfnamefont {Q.}~\bibnamefont {Si}},\ }\href {\doibase
  10.1073/pnas.1715851115} {\bibfield  {journal} {\bibinfo  {journal} {Proc.
  Natl. Acad. Sci. U.S.A.}\ }\textbf {\bibinfo {volume} {115}},\ \bibinfo
  {pages} {93} (\bibinfo {year} {2018})}\BibitemShut {NoStop}%
\bibitem [{\citenamefont {Dzsaber}\ \emph {et~al.}(2017)\citenamefont
  {Dzsaber}, \citenamefont {Prochaska}, \citenamefont {Sidorenko},
  \citenamefont {Eguchi}, \citenamefont {Svagera}, \citenamefont {Waas},
  \citenamefont {Prokofiev}, \citenamefont {Si},\ and\ \citenamefont
  {Paschen}}]{Dzsaber2017}%
  \BibitemOpen
  \bibfield  {author} {\bibinfo {author} {\bibfnamefont {S.}~\bibnamefont
  {Dzsaber}}, \bibinfo {author} {\bibfnamefont {L.}~\bibnamefont {Prochaska}},
  \bibinfo {author} {\bibfnamefont {A.}~\bibnamefont {Sidorenko}}, \bibinfo
  {author} {\bibfnamefont {G.}~\bibnamefont {Eguchi}}, \bibinfo {author}
  {\bibfnamefont {R.}~\bibnamefont {Svagera}}, \bibinfo {author} {\bibfnamefont
  {M.}~\bibnamefont {Waas}}, \bibinfo {author} {\bibfnamefont {A.}~\bibnamefont
  {Prokofiev}}, \bibinfo {author} {\bibfnamefont {Q.}~\bibnamefont {Si}}, \
  and\ \bibinfo {author} {\bibfnamefont {S.}~\bibnamefont {Paschen}},\ }\href
  {\doibase 10.1103/PhysRevLett.118.246601} {\bibfield  {journal} {\bibinfo
  {journal} {Phys. Rev. Lett.}\ }\textbf {\bibinfo {volume} {118}},\ \bibinfo
  {pages} {246601} (\bibinfo {year} {2017})}\BibitemShut {NoStop}%
\bibitem [{\citenamefont {Dzsaber}\ \emph {et~al.}(2021)\citenamefont
  {Dzsaber}, \citenamefont {Yan}, \citenamefont {Taupin}, \citenamefont
  {Eguchi}, \citenamefont {Prokofiev}, \citenamefont {Shiroka}, \citenamefont
  {Blaha}, \citenamefont {Rubel}, \citenamefont {Grefe}, \citenamefont {Lai},
  \citenamefont {Si},\ and\ \citenamefont {Paschen}}]{Dzs-giant21.1}%
  \BibitemOpen
  \bibfield  {author} {\bibinfo {author} {\bibfnamefont {S.}~\bibnamefont
  {Dzsaber}}, \bibinfo {author} {\bibfnamefont {X.}~\bibnamefont {Yan}},
  \bibinfo {author} {\bibfnamefont {M.}~\bibnamefont {Taupin}}, \bibinfo
  {author} {\bibfnamefont {G.}~\bibnamefont {Eguchi}}, \bibinfo {author}
  {\bibfnamefont {A.}~\bibnamefont {Prokofiev}}, \bibinfo {author}
  {\bibfnamefont {T.}~\bibnamefont {Shiroka}}, \bibinfo {author} {\bibfnamefont
  {P.}~\bibnamefont {Blaha}}, \bibinfo {author} {\bibfnamefont
  {O.}~\bibnamefont {Rubel}}, \bibinfo {author} {\bibfnamefont {S.~E.}\
  \bibnamefont {Grefe}}, \bibinfo {author} {\bibfnamefont {H.-H.}\ \bibnamefont
  {Lai}}, \bibinfo {author} {\bibfnamefont {Q.}~\bibnamefont {Si}}, \ and\
  \bibinfo {author} {\bibfnamefont {S.}~\bibnamefont {Paschen}},\ }\href
  {\doibase 10.1073/pnas.2013386118} {\bibfield  {journal} {\bibinfo  {journal}
  {Proc. Natl. Acad. Sci. U.S.A.}\ }\textbf {\bibinfo {volume} {118}},\
  \bibinfo {pages} {e2013386118} (\bibinfo {year} {2021})}\BibitemShut
  {NoStop}%
\bibitem [{\citenamefont {Chen}\ \emph {et~al.}(2022)\citenamefont {Chen},
  \citenamefont {Setty}, \citenamefont {Hu}, \citenamefont {Vergniory},
  \citenamefont {Grefe}, \citenamefont {Fischer}, \citenamefont {Yan},
  \citenamefont {Eguchi}, \citenamefont {Prokofiev}, \citenamefont {Paschen},
  \citenamefont {Cano},\ and\ \citenamefont {Si}}]{Chen-Si2022}%
  \BibitemOpen
  \bibfield  {author} {\bibinfo {author} {\bibfnamefont {L.}~\bibnamefont
  {Chen}}, \bibinfo {author} {\bibfnamefont {C.}~\bibnamefont {Setty}},
  \bibinfo {author} {\bibfnamefont {H.}~\bibnamefont {Hu}}, \bibinfo {author}
  {\bibfnamefont {M.~G.}\ \bibnamefont {Vergniory}}, \bibinfo {author}
  {\bibfnamefont {S.~E.}\ \bibnamefont {Grefe}}, \bibinfo {author}
  {\bibfnamefont {L.}~\bibnamefont {Fischer}}, \bibinfo {author} {\bibfnamefont
  {X.}~\bibnamefont {Yan}}, \bibinfo {author} {\bibfnamefont {G.}~\bibnamefont
  {Eguchi}}, \bibinfo {author} {\bibfnamefont {A.}~\bibnamefont {Prokofiev}},
  \bibinfo {author} {\bibfnamefont {S.}~\bibnamefont {Paschen}}, \bibinfo
  {author} {\bibfnamefont {J.}~\bibnamefont {Cano}}, \ and\ \bibinfo {author}
  {\bibfnamefont {Q.}~\bibnamefont {Si}},\ }\href {\doibase
  10.1038/s41567-022-01743-4} {\bibfield  {journal} {\bibinfo  {journal}
  {{Nat.\ Phys.}}\ }\textbf {\bibinfo {volume} {18}},\ \bibinfo {pages}
  {134101346} (\bibinfo {year} {2022})}\BibitemShut {NoStop}%
\bibitem [{\citenamefont {Armitage}\ \emph {et~al.}(2018)\citenamefont
  {Armitage}, \citenamefont {Mele},\ and\ \citenamefont
  {Vishwanath}}]{Armitage2017}%
  \BibitemOpen
  \bibfield  {author} {\bibinfo {author} {\bibfnamefont {N.~P.}\ \bibnamefont
  {Armitage}}, \bibinfo {author} {\bibfnamefont {E.~J.}\ \bibnamefont {Mele}},
  \ and\ \bibinfo {author} {\bibfnamefont {A.}~\bibnamefont {Vishwanath}},\
  }\href {\doibase 10.1103/RevModPhys.90.015001} {\bibfield  {journal}
  {\bibinfo  {journal} {Rev. Mod. Phys.}\ }\textbf {\bibinfo {volume} {90}},\
  \bibinfo {pages} {015001} (\bibinfo {year} {2018})}\BibitemShut {NoStop}%
\bibitem [{\citenamefont {Nagaosa}\ \emph {et~al.}(2020)\citenamefont
  {Nagaosa}, \citenamefont {Morimoto},\ and\ \citenamefont
  {Tokura}}]{Nagaosa2020}%
  \BibitemOpen
  \bibfield  {author} {\bibinfo {author} {\bibfnamefont {N.}~\bibnamefont
  {Nagaosa}}, \bibinfo {author} {\bibfnamefont {T.}~\bibnamefont {Morimoto}}, \
  and\ \bibinfo {author} {\bibfnamefont {Y.}~\bibnamefont {Tokura}},\ }\href
  {\doibase 10.1038/s41578-020-0208-y} {\bibfield  {journal} {\bibinfo
  {journal} {Nature Reviews Materials}\ }\textbf {\bibinfo {volume} {5}},\
  \bibinfo {pages} {621} (\bibinfo {year} {2020})}\BibitemShut {NoStop}%
\bibitem [{\citenamefont {Bradlyn}\ \emph {et~al.}(2017)\citenamefont
  {Bradlyn}, \citenamefont {Elcoro}, \citenamefont {Cano}, \citenamefont
  {Vergniory}, \citenamefont {Wang}, \citenamefont {Felser}, \citenamefont
  {Aroyo},\ and\ \citenamefont {Bernevig}}]{Bradlyn2017}%
  \BibitemOpen
  \bibfield  {author} {\bibinfo {author} {\bibfnamefont {B.}~\bibnamefont
  {Bradlyn}}, \bibinfo {author} {\bibfnamefont {L.}~\bibnamefont {Elcoro}},
  \bibinfo {author} {\bibfnamefont {J.}~\bibnamefont {Cano}}, \bibinfo {author}
  {\bibfnamefont {M.~G.}\ \bibnamefont {Vergniory}}, \bibinfo {author}
  {\bibfnamefont {Z.}~\bibnamefont {Wang}}, \bibinfo {author} {\bibfnamefont
  {C.}~\bibnamefont {Felser}}, \bibinfo {author} {\bibfnamefont {M.~I.}\
  \bibnamefont {Aroyo}}, \ and\ \bibinfo {author} {\bibfnamefont {B.~A.}\
  \bibnamefont {Bernevig}},\ }\href {\doibase 10.1038/nature23268} {\bibfield
  {journal} {\bibinfo  {journal} {Nature}\ }\textbf {\bibinfo {volume} {547}},\
  \bibinfo {pages} {298} (\bibinfo {year} {2017})}\BibitemShut {NoStop}%
\bibitem [{\citenamefont {Cano}\ \emph {et~al.}(2018)\citenamefont {Cano},
  \citenamefont {Bradlyn}, \citenamefont {Wang}, \citenamefont {Elcoro},
  \citenamefont {Vergniory}, \citenamefont {Felser}, \citenamefont {Aroyo},\
  and\ \citenamefont {Bernevig}}]{Cano2018}%
  \BibitemOpen
  \bibfield  {author} {\bibinfo {author} {\bibfnamefont {J.}~\bibnamefont
  {Cano}}, \bibinfo {author} {\bibfnamefont {B.}~\bibnamefont {Bradlyn}},
  \bibinfo {author} {\bibfnamefont {Z.}~\bibnamefont {Wang}}, \bibinfo {author}
  {\bibfnamefont {L.}~\bibnamefont {Elcoro}}, \bibinfo {author} {\bibfnamefont
  {M.~G.}\ \bibnamefont {Vergniory}}, \bibinfo {author} {\bibfnamefont
  {C.}~\bibnamefont {Felser}}, \bibinfo {author} {\bibfnamefont {M.~I.}\
  \bibnamefont {Aroyo}}, \ and\ \bibinfo {author} {\bibfnamefont {B.~A.}\
  \bibnamefont {Bernevig}},\ }\href {\doibase 10.1103/PhysRevB.97.035139}
  {\bibfield  {journal} {\bibinfo  {journal} {Phys. Rev. B}\ }\textbf {\bibinfo
  {volume} {97}},\ \bibinfo {pages} {035139} (\bibinfo {year}
  {2018})}\BibitemShut {NoStop}%
\bibitem [{\citenamefont {Po}\ \emph {et~al.}(2017)\citenamefont {Po},
  \citenamefont {Vishwanath},\ and\ \citenamefont {Watanabe}}]{Po2017}%
  \BibitemOpen
  \bibfield  {author} {\bibinfo {author} {\bibfnamefont {H.~C.}\ \bibnamefont
  {Po}}, \bibinfo {author} {\bibfnamefont {A.}~\bibnamefont {Vishwanath}}, \
  and\ \bibinfo {author} {\bibfnamefont {H.}~\bibnamefont {Watanabe}},\ }\href
  {\doibase 10.1038/s41467-017-00133-2} {\bibfield  {journal} {\bibinfo
  {journal} {Nature Communications}\ }\textbf {\bibinfo {volume} {8}},\
  \bibinfo {pages} {50} (\bibinfo {year} {2017})}\BibitemShut {NoStop}%
\bibitem [{\citenamefont {Watanabe}\ \emph {et~al.}(2016)\citenamefont
  {Watanabe}, \citenamefont {Po}, \citenamefont {Zaletel},\ and\ \citenamefont
  {Vishwanath}}]{Watanabe2017}%
  \BibitemOpen
  \bibfield  {author} {\bibinfo {author} {\bibfnamefont {H.}~\bibnamefont
  {Watanabe}}, \bibinfo {author} {\bibfnamefont {H.~C.}\ \bibnamefont {Po}},
  \bibinfo {author} {\bibfnamefont {M.~P.}\ \bibnamefont {Zaletel}}, \ and\
  \bibinfo {author} {\bibfnamefont {A.}~\bibnamefont {Vishwanath}},\ }\href
  {\doibase 10.1103/PhysRevLett.117.096404} {\bibfield  {journal} {\bibinfo
  {journal} {Phys. Rev. Lett.}\ }\textbf {\bibinfo {volume} {117}},\ \bibinfo
  {pages} {096404} (\bibinfo {year} {2016})}\BibitemShut {NoStop}%
\bibitem [{\citenamefont {Cano}\ and\ \citenamefont
  {Bradlyn}(2021)}]{cano2021band}%
  \BibitemOpen
  \bibfield  {author} {\bibinfo {author} {\bibfnamefont {J.}~\bibnamefont
  {Cano}}\ and\ \bibinfo {author} {\bibfnamefont {B.}~\bibnamefont {Bradlyn}},\
  }\href@noop {} {\bibfield  {journal} {\bibinfo  {journal} {Annu.\ Rev.\
  Condens.\ Matter Phys.}\ }\textbf {\bibinfo {volume} {12}},\ \bibinfo {pages}
  {225} (\bibinfo {year} {2021})}\BibitemShut {NoStop}%
\bibitem [{\citenamefont {Abrikosov}\ \emph {et~al.}(2012)\citenamefont
  {Abrikosov}, \citenamefont {Gorkov},\ and\ \citenamefont
  {Dzyaloshinski}}]{AGD}%
  \BibitemOpen
  \bibfield  {author} {\bibinfo {author} {\bibfnamefont {A.~A.}\ \bibnamefont
  {Abrikosov}}, \bibinfo {author} {\bibfnamefont {L.~P.}\ \bibnamefont
  {Gorkov}}, \ and\ \bibinfo {author} {\bibfnamefont {I.~E.}\ \bibnamefont
  {Dzyaloshinski}},\ }\href@noop {} {\emph {\bibinfo {title} {Methods of
  quantum field theory in statistical physics}}}\ (\bibinfo  {publisher}
  {Courier Corporation},\ \bibinfo {year} {2012})\BibitemShut {NoStop}%
\bibitem [{\citenamefont {Hu}\ \emph {et~al.}(2021)\citenamefont {Hu},
  \citenamefont {Chen}, \citenamefont {Setty}, \citenamefont {Grefe},
  \citenamefont {Prokofiev}, \citenamefont {Kirchner}, \citenamefont {Paschen},
  \citenamefont {Cano},\ and\ \citenamefont {Si}}]{Hu-Si2021}%
  \BibitemOpen
  \bibfield  {author} {\bibinfo {author} {\bibfnamefont {H.}~\bibnamefont
  {Hu}}, \bibinfo {author} {\bibfnamefont {L.}~\bibnamefont {Chen}}, \bibinfo
  {author} {\bibfnamefont {C.}~\bibnamefont {Setty}}, \bibinfo {author}
  {\bibfnamefont {S.~E.}\ \bibnamefont {Grefe}}, \bibinfo {author}
  {\bibfnamefont {A.}~\bibnamefont {Prokofiev}}, \bibinfo {author}
  {\bibfnamefont {S.}~\bibnamefont {Kirchner}}, \bibinfo {author}
  {\bibfnamefont {S.}~\bibnamefont {Paschen}}, \bibinfo {author} {\bibfnamefont
  {J.}~\bibnamefont {Cano}}, \ and\ \bibinfo {author} {\bibfnamefont
  {Q.}~\bibnamefont {Si}},\ }\href@noop {} {\bibfield  {journal} {\bibinfo
  {journal} {arXiv preprint arXiv:2110.06182}\ } (\bibinfo {year}
  {2021})}\BibitemShut {NoStop}%
\bibitem [{\citenamefont {Setty}\ \emph
  {et~al.}(2023{\natexlab{a}})\citenamefont {Setty}, \citenamefont {Xie},
  \citenamefont {Sur}, \citenamefont {Chen}, \citenamefont {Paschen},
  \citenamefont {Vergniory}, \citenamefont {Cano},\ and\ \citenamefont
  {Si}}]{Setty_v3}%
  \BibitemOpen
  \bibfield  {author} {\bibinfo {author} {\bibfnamefont {C.}~\bibnamefont
  {Setty}}, \bibinfo {author} {\bibfnamefont {F.}~\bibnamefont {Xie}}, \bibinfo
  {author} {\bibfnamefont {S.}~\bibnamefont {Sur}}, \bibinfo {author}
  {\bibfnamefont {L.}~\bibnamefont {Chen}}, \bibinfo {author} {\bibfnamefont
  {S.}~\bibnamefont {Paschen}}, \bibinfo {author} {\bibfnamefont {M.~G.}\
  \bibnamefont {Vergniory}}, \bibinfo {author} {\bibfnamefont {J.}~\bibnamefont
  {Cano}}, \ and\ \bibinfo {author} {\bibfnamefont {Q.}~\bibnamefont {Si}},\
  }\href@noop {} {\bibfield  {journal} {\bibinfo  {journal} {arXiv preprint
  arXiv:2311.12031}\ } (\bibinfo {year} {2023}{\natexlab{a}})}\BibitemShut
  {NoStop}%
\bibitem [{\citenamefont {Dzyaloshinskii}(2003)}]{Dzyaloshinskii2003}%
  \BibitemOpen
  \bibfield  {author} {\bibinfo {author} {\bibfnamefont {I.}~\bibnamefont
  {Dzyaloshinskii}},\ }\href {\doibase 10.1103/PhysRevB.68.085113} {\bibfield
  {journal} {\bibinfo  {journal} {Phys. Rev. B}\ }\textbf {\bibinfo {volume}
  {68}},\ \bibinfo {pages} {085113} (\bibinfo {year} {2003})}\BibitemShut
  {NoStop}%
\bibitem [{\citenamefont {Essin}\ and\ \citenamefont
  {Gurarie}(2011)}]{Gurarie2011-2}%
  \BibitemOpen
  \bibfield  {author} {\bibinfo {author} {\bibfnamefont {A.~M.}\ \bibnamefont
  {Essin}}\ and\ \bibinfo {author} {\bibfnamefont {V.}~\bibnamefont
  {Gurarie}},\ }\href {\doibase 10.1103/PhysRevB.84.125132} {\bibfield
  {journal} {\bibinfo  {journal} {Phys. Rev. B}\ }\textbf {\bibinfo {volume}
  {84}},\ \bibinfo {pages} {125132} (\bibinfo {year} {2011})}\BibitemShut
  {NoStop}%
\bibitem [{\citenamefont {You}\ \emph {et~al.}(2018)\citenamefont {You},
  \citenamefont {He}, \citenamefont {Xu},\ and\ \citenamefont
  {Vishwanath}}]{You-smg2018}%
  \BibitemOpen
  \bibfield  {author} {\bibinfo {author} {\bibfnamefont {Y.-Z.}\ \bibnamefont
  {You}}, \bibinfo {author} {\bibfnamefont {Y.-C.}\ \bibnamefont {He}},
  \bibinfo {author} {\bibfnamefont {C.}~\bibnamefont {Xu}}, \ and\ \bibinfo
  {author} {\bibfnamefont {A.}~\bibnamefont {Vishwanath}},\ }\href {\doibase
  10.1103/PhysRevX.8.011026} {\bibfield  {journal} {\bibinfo  {journal} {Phys.
  Rev. X}\ }\textbf {\bibinfo {volume} {8}},\ \bibinfo {pages} {011026}
  (\bibinfo {year} {2018})}\BibitemShut {NoStop}%
\bibitem [{\citenamefont {Setty}\ \emph
  {et~al.}(2023{\natexlab{b}})\citenamefont {Setty}, \citenamefont {Sur},
  \citenamefont {Chen}, \citenamefont {Xie}, \citenamefont {Hu}, \citenamefont
  {Paschen}, \citenamefont {Cano},\ and\ \citenamefont {Si}}]{Setty_v1}%
  \BibitemOpen
  \bibfield  {author} {\bibinfo {author} {\bibfnamefont {C.}~\bibnamefont
  {Setty}}, \bibinfo {author} {\bibfnamefont {S.}~\bibnamefont {Sur}}, \bibinfo
  {author} {\bibfnamefont {L.}~\bibnamefont {Chen}}, \bibinfo {author}
  {\bibfnamefont {F.}~\bibnamefont {Xie}}, \bibinfo {author} {\bibfnamefont
  {H.}~\bibnamefont {Hu}}, \bibinfo {author} {\bibfnamefont {S.}~\bibnamefont
  {Paschen}}, \bibinfo {author} {\bibfnamefont {J.}~\bibnamefont {Cano}}, \
  and\ \bibinfo {author} {\bibfnamefont {Q.}~\bibnamefont {Si}},\ }\href@noop
  {} {\bibfield  {journal} {\bibinfo  {journal} {arXiv preprint
  arXiv:2301.13870}\ } (\bibinfo {year} {2023}{\natexlab{b}})}\BibitemShut
  {NoStop}%
\bibitem [{\citenamefont {Wagner}\ \emph
  {et~al.}(2023{\natexlab{a}})\citenamefont {Wagner}, \citenamefont {Crippa},
  \citenamefont {Amaricci}, \citenamefont {Hansmann}, \citenamefont {Klett},
  \citenamefont {K{\"o}nig}, \citenamefont {Sch{\"a}fer}, \citenamefont
  {Sante}, \citenamefont {Cano}, \citenamefont {Millis}, \citenamefont
  {Georges},\ and\ \citenamefont {Sangiovanni}}]{Sangiovanni_1}%
  \BibitemOpen
  \bibfield  {author} {\bibinfo {author} {\bibfnamefont {N.}~\bibnamefont
  {Wagner}}, \bibinfo {author} {\bibfnamefont {L.}~\bibnamefont {Crippa}},
  \bibinfo {author} {\bibfnamefont {A.}~\bibnamefont {Amaricci}}, \bibinfo
  {author} {\bibfnamefont {P.}~\bibnamefont {Hansmann}}, \bibinfo {author}
  {\bibfnamefont {M.}~\bibnamefont {Klett}}, \bibinfo {author} {\bibfnamefont
  {E.~J.}\ \bibnamefont {K{\"o}nig}}, \bibinfo {author} {\bibfnamefont
  {T.}~\bibnamefont {Sch{\"a}fer}}, \bibinfo {author} {\bibfnamefont {D.~D.}\
  \bibnamefont {Sante}}, \bibinfo {author} {\bibfnamefont {J.}~\bibnamefont
  {Cano}}, \bibinfo {author} {\bibfnamefont {A.~J.}\ \bibnamefont {Millis}},
  \bibinfo {author} {\bibfnamefont {A.}~\bibnamefont {Georges}}, \ and\
  \bibinfo {author} {\bibfnamefont {G.}~\bibnamefont {Sangiovanni}},\ }\href
  {\doibase 10.1038/s41467-023-42773-7} {\bibfield  {journal} {\bibinfo
  {journal} {Nature Communications}\ }\textbf {\bibinfo {volume} {14}},\
  \bibinfo {pages} {7531} (\bibinfo {year} {2023}{\natexlab{a}})}\BibitemShut
  {NoStop}%
\bibitem [{\citenamefont {Morimoto}\ and\ \citenamefont
  {Nagaosa}(2016)}]{Morimoto2016}%
  \BibitemOpen
  \bibfield  {author} {\bibinfo {author} {\bibfnamefont {T.}~\bibnamefont
  {Morimoto}}\ and\ \bibinfo {author} {\bibfnamefont {N.}~\bibnamefont
  {Nagaosa}},\ }\href {\doibase 10.1038/srep19853} {\bibfield  {journal}
  {\bibinfo  {journal} {Sci. Rep.}\ }\textbf {\bibinfo {volume} {6}},\ \bibinfo
  {pages} {19853} (\bibinfo {year} {2016})}\BibitemShut {NoStop}%
\bibitem [{\citenamefont {Kirchner}\ \emph {et~al.}(2020)\citenamefont
  {Kirchner}, \citenamefont {Paschen}, \citenamefont {Chen}, \citenamefont
  {Wirth}, \citenamefont {Feng}, \citenamefont {Thompson},\ and\ \citenamefont
  {Si}}]{Kir20.1}%
  \BibitemOpen
  \bibfield  {author} {\bibinfo {author} {\bibfnamefont {S.}~\bibnamefont
  {Kirchner}}, \bibinfo {author} {\bibfnamefont {S.}~\bibnamefont {Paschen}},
  \bibinfo {author} {\bibfnamefont {Q.}~\bibnamefont {Chen}}, \bibinfo {author}
  {\bibfnamefont {S.}~\bibnamefont {Wirth}}, \bibinfo {author} {\bibfnamefont
  {D.}~\bibnamefont {Feng}}, \bibinfo {author} {\bibfnamefont {J.~D.}\
  \bibnamefont {Thompson}}, \ and\ \bibinfo {author} {\bibfnamefont
  {Q.}~\bibnamefont {Si}},\ }\href {\doibase 10.1103/RevModPhys.92.011002}
  {\bibfield  {journal} {\bibinfo  {journal} {Rev. Mod. Phys.}\ }\textbf
  {\bibinfo {volume} {92}},\ \bibinfo {pages} {011002} (\bibinfo {year}
  {2020})}\BibitemShut {NoStop}%
\bibitem [{\citenamefont {Si}\ \emph {et~al.}(2001)\citenamefont {Si},
  \citenamefont {Rabello}, \citenamefont {Ingersent},\ and\ \citenamefont
  {Smith}}]{Si01.1}%
  \BibitemOpen
  \bibfield  {author} {\bibinfo {author} {\bibfnamefont {Q.}~\bibnamefont
  {Si}}, \bibinfo {author} {\bibfnamefont {S.}~\bibnamefont {Rabello}},
  \bibinfo {author} {\bibfnamefont {K.}~\bibnamefont {Ingersent}}, \ and\
  \bibinfo {author} {\bibfnamefont {J.}~\bibnamefont {Smith}},\ }\href
  {\doibase 10.1038/35101507} {\bibfield  {journal} {\bibinfo  {journal}
  {Nature}\ }\textbf {\bibinfo {volume} {413}},\ \bibinfo {pages} {804}
  (\bibinfo {year} {2001})}\BibitemShut {NoStop}%
\bibitem [{\citenamefont {Coleman}\ \emph {et~al.}(2001)\citenamefont
  {Coleman}, \citenamefont {P\'epin}, \citenamefont {Si},\ and\ \citenamefont
  {Ramazashvili}}]{Col01.1}%
  \BibitemOpen
  \bibfield  {author} {\bibinfo {author} {\bibfnamefont {P.}~\bibnamefont
  {Coleman}}, \bibinfo {author} {\bibfnamefont {C.}~\bibnamefont {P\'epin}},
  \bibinfo {author} {\bibfnamefont {Q.}~\bibnamefont {Si}}, \ and\ \bibinfo
  {author} {\bibfnamefont {R.}~\bibnamefont {Ramazashvili}},\ }\href {\doibase
  10.1088/0953-8984/13/35/202} {\bibfield  {journal} {\bibinfo  {journal} {{J.\
  Phys.: Condens.\ Matter}}\ }\textbf {\bibinfo {volume} {13}},\ \bibinfo
  {pages} {R723} (\bibinfo {year} {2001})}\BibitemShut {NoStop}%
\bibitem [{\citenamefont {Senthil}\ \emph {et~al.}(2004)\citenamefont
  {Senthil}, \citenamefont {Vojta},\ and\ \citenamefont {Sachdev}}]{Sen04.1}%
  \BibitemOpen
  \bibfield  {author} {\bibinfo {author} {\bibfnamefont {T.}~\bibnamefont
  {Senthil}}, \bibinfo {author} {\bibfnamefont {M.}~\bibnamefont {Vojta}}, \
  and\ \bibinfo {author} {\bibfnamefont {S.}~\bibnamefont {Sachdev}},\ }\href
  {\doibase 10.1103/PhysRevB.69.035111} {\bibfield  {journal} {\bibinfo
  {journal} {Phys. Rev. B}\ }\textbf {\bibinfo {volume} {69}},\ \bibinfo
  {pages} {035111} (\bibinfo {year} {2004})}\BibitemShut {NoStop}%
\bibitem [{\citenamefont {Anisimov}\ \emph {et~al.}(2002)\citenamefont
  {Anisimov}, \citenamefont {Nekrasov}, \citenamefont {Kondakov}, \citenamefont
  {Rice},\ and\ \citenamefont {Sigrist}}]{Anisimov2002}%
  \BibitemOpen
  \bibfield  {author} {\bibinfo {author} {\bibfnamefont {V.}~\bibnamefont
  {Anisimov}}, \bibinfo {author} {\bibfnamefont {I.}~\bibnamefont {Nekrasov}},
  \bibinfo {author} {\bibfnamefont {D.}~\bibnamefont {Kondakov}}, \bibinfo
  {author} {\bibfnamefont {T.}~\bibnamefont {Rice}}, \ and\ \bibinfo {author}
  {\bibfnamefont {M.}~\bibnamefont {Sigrist}},\ }\href@noop {} {\bibfield
  {journal} {\bibinfo  {journal} {The European Physical Journal B-Condensed
  Matter and Complex Systems}\ }\textbf {\bibinfo {volume} {25}},\ \bibinfo
  {pages} {191} (\bibinfo {year} {2002})}\BibitemShut {NoStop}%
\bibitem [{\citenamefont {Yi}\ \emph {et~al.}(2017)\citenamefont {Yi},
  \citenamefont {Zhang}, \citenamefont {Shen},\ and\ \citenamefont
  {Lu}}]{Yi17.1}%
  \BibitemOpen
  \bibfield  {author} {\bibinfo {author} {\bibfnamefont {M.}~\bibnamefont
  {Yi}}, \bibinfo {author} {\bibfnamefont {Y.}~\bibnamefont {Zhang}}, \bibinfo
  {author} {\bibfnamefont {Z.-X.}\ \bibnamefont {Shen}}, \ and\ \bibinfo
  {author} {\bibfnamefont {D.}~\bibnamefont {Lu}},\ }\href {\doibase
  10.1038/s41535-017-0059-y} {\bibfield  {journal} {\bibinfo  {journal} {{npj
  Quantum Mater.}}\ }\textbf {\bibinfo {volume} {2}},\ \bibinfo {pages} {57}
  (\bibinfo {year} {2017})}\BibitemShut {NoStop}%
\bibitem [{\citenamefont {Si}\ and\ \citenamefont
  {Hussey}(2023)}]{Si-Hussey23}%
  \BibitemOpen
  \bibfield  {author} {\bibinfo {author} {\bibfnamefont {Q.}~\bibnamefont
  {Si}}\ and\ \bibinfo {author} {\bibfnamefont {N.~E.}\ \bibnamefont
  {Hussey}},\ }\href {\doibase 10.1063/PT.3.5235} {\bibfield  {journal}
  {\bibinfo  {journal} {{Phys.\ Today}}\ }\textbf {\bibinfo {volume} {76}},\
  \bibinfo {pages} {34} (\bibinfo {year} {2023})}\BibitemShut {NoStop}%
\bibitem [{\citenamefont {Zhao}\ \emph
  {et~al.}(2023{\natexlab{a}})\citenamefont {Zhao}, \citenamefont {Shen},
  \citenamefont {Tao}, \citenamefont {Han}, \citenamefont {Kang}, \citenamefont
  {Watanabe}, \citenamefont {Taniguchi}, \citenamefont {Mak},\ and\
  \citenamefont {Shan}}]{Shan2023}%
  \BibitemOpen
  \bibfield  {author} {\bibinfo {author} {\bibfnamefont {W.}~\bibnamefont
  {Zhao}}, \bibinfo {author} {\bibfnamefont {B.}~\bibnamefont {Shen}}, \bibinfo
  {author} {\bibfnamefont {Z.}~\bibnamefont {Tao}}, \bibinfo {author}
  {\bibfnamefont {Z.}~\bibnamefont {Han}}, \bibinfo {author} {\bibfnamefont
  {K.}~\bibnamefont {Kang}}, \bibinfo {author} {\bibfnamefont {K.}~\bibnamefont
  {Watanabe}}, \bibinfo {author} {\bibfnamefont {T.}~\bibnamefont {Taniguchi}},
  \bibinfo {author} {\bibfnamefont {K.~F.}\ \bibnamefont {Mak}}, \ and\
  \bibinfo {author} {\bibfnamefont {J.}~\bibnamefont {Shan}},\ }\href {\doibase
  10.1038/s41586-023-05800-7} {\bibfield  {journal} {\bibinfo  {journal}
  {Nature}\ }\textbf {\bibinfo {volume} {616}},\ \bibinfo {pages} {61}
  (\bibinfo {year} {2023}{\natexlab{a}})}\BibitemShut {NoStop}%
\bibitem [{\citenamefont {Guerci}\ \emph {et~al.}(2023)\citenamefont {Guerci},
  \citenamefont {Wang}, \citenamefont {Zang}, \citenamefont {Cano},
  \citenamefont {Pixley},\ and\ \citenamefont {Millis}}]{Guerci2022.x}%
  \BibitemOpen
  \bibfield  {author} {\bibinfo {author} {\bibfnamefont {D.}~\bibnamefont
  {Guerci}}, \bibinfo {author} {\bibfnamefont {J.}~\bibnamefont {Wang}},
  \bibinfo {author} {\bibfnamefont {J.}~\bibnamefont {Zang}}, \bibinfo {author}
  {\bibfnamefont {J.}~\bibnamefont {Cano}}, \bibinfo {author} {\bibfnamefont
  {J.~H.}\ \bibnamefont {Pixley}}, \ and\ \bibinfo {author} {\bibfnamefont
  {A.}~\bibnamefont {Millis}},\ }\href {\doibase 10.1126/sciadv.ade7701}
  {\bibfield  {journal} {\bibinfo  {journal} {Sci. Adv.}\ }\textbf {\bibinfo
  {volume} {9}},\ \bibinfo {pages} {eade7701} (\bibinfo {year}
  {2023})}\BibitemShut {NoStop}%
\bibitem [{\citenamefont {Xie}\ \emph {et~al.}(2023)\citenamefont {Xie},
  \citenamefont {Chen},\ and\ \citenamefont {Si}}]{Xie2023.x}%
  \BibitemOpen
  \bibfield  {author} {\bibinfo {author} {\bibfnamefont {F.}~\bibnamefont
  {Xie}}, \bibinfo {author} {\bibfnamefont {L.}~\bibnamefont {Chen}}, \ and\
  \bibinfo {author} {\bibfnamefont {Q.}~\bibnamefont {Si}},\ }\href@noop {}
  {\bibfield  {journal} {\bibinfo  {journal} {arXiv preprint arXiv:2310.20676}\
  } (\bibinfo {year} {2023})}\BibitemShut {NoStop}%
\bibitem [{\citenamefont {Zhao}\ \emph
  {et~al.}(2023{\natexlab{b}})\citenamefont {Zhao}, \citenamefont {Shen},
  \citenamefont {Tao}, \citenamefont {Kim}, \citenamefont {Kn{\"u}ppel},
  \citenamefont {Han}, \citenamefont {Zhang}, \citenamefont {Watanabe},
  \citenamefont {Taniguchi}, \citenamefont {Chowdhury}, \citenamefont {Shan},\
  and\ \citenamefont {Mak}}]{Zhao2023.x}%
  \BibitemOpen
  \bibfield  {author} {\bibinfo {author} {\bibfnamefont {W.}~\bibnamefont
  {Zhao}}, \bibinfo {author} {\bibfnamefont {B.}~\bibnamefont {Shen}}, \bibinfo
  {author} {\bibfnamefont {Z.}~\bibnamefont {Tao}}, \bibinfo {author}
  {\bibfnamefont {S.}~\bibnamefont {Kim}}, \bibinfo {author} {\bibfnamefont
  {P.}~\bibnamefont {Kn{\"u}ppel}}, \bibinfo {author} {\bibfnamefont
  {Z.}~\bibnamefont {Han}}, \bibinfo {author} {\bibfnamefont {Y.}~\bibnamefont
  {Zhang}}, \bibinfo {author} {\bibfnamefont {K.}~\bibnamefont {Watanabe}},
  \bibinfo {author} {\bibfnamefont {T.}~\bibnamefont {Taniguchi}}, \bibinfo
  {author} {\bibfnamefont {D.}~\bibnamefont {Chowdhury}}, \bibinfo {author}
  {\bibfnamefont {J.}~\bibnamefont {Shan}}, \ and\ \bibinfo {author}
  {\bibfnamefont {K.~F.}\ \bibnamefont {Mak}},\ }\href@noop {} {\bibfield
  {journal} {\bibinfo  {journal} {arXiv preprint arXiv:2310.06044}\ } (\bibinfo
  {year} {2023}{\natexlab{b}})}\BibitemShut {NoStop}%
\bibitem [{\citenamefont {Hu}\ and\ \citenamefont {Si}(2023)}]{Hu-coupled22.3}%
  \BibitemOpen
  \bibfield  {author} {\bibinfo {author} {\bibfnamefont {H.}~\bibnamefont
  {Hu}}\ and\ \bibinfo {author} {\bibfnamefont {Q.}~\bibnamefont {Si}},\ }\href
  {\doibase 10.1126/sciadv.adg0028} {\bibfield  {journal} {\bibinfo  {journal}
  {Sci. Adv.}\ }\textbf {\bibinfo {volume} {9}},\ \bibinfo {pages} {eadg0028}
  (\bibinfo {year} {2023})}\BibitemShut {NoStop}%
\bibitem [{\citenamefont {Huang}\ \emph {et~al.}()\citenamefont {Huang},
  \citenamefont {Setty}, \citenamefont {Deng}, \citenamefont {You},
  \citenamefont {Liu}, \citenamefont {Shao}, \citenamefont {Oh}, \citenamefont
  {Guo}, \citenamefont {Zhang}, \citenamefont {Yue}, \citenamefont {Yin},
  \citenamefont {Hashimoto}, \citenamefont {Lu}, \citenamefont {Gorovikov},
  \citenamefont {Dai}, \citenamefont {Hasan}, \citenamefont {Feng},
  \citenamefont {Birgeneau}, \citenamefont {Shi}, \citenamefont {Chu},
  \citenamefont {Chang}, \citenamefont {Si},\ and\ \citenamefont
  {Yi}}]{Hua23.3x}%
  \BibitemOpen
  \bibfield  {author} {\bibinfo {author} {\bibfnamefont {J.}~\bibnamefont
  {Huang}}, \bibinfo {author} {\bibfnamefont {C.}~\bibnamefont {Setty}},
  \bibinfo {author} {\bibfnamefont {L.}~\bibnamefont {Deng}}, \bibinfo {author}
  {\bibfnamefont {J.-Y.}\ \bibnamefont {You}}, \bibinfo {author} {\bibfnamefont
  {H.}~\bibnamefont {Liu}}, \bibinfo {author} {\bibfnamefont {S.}~\bibnamefont
  {Shao}}, \bibinfo {author} {\bibfnamefont {J.~S.}\ \bibnamefont {Oh}},
  \bibinfo {author} {\bibfnamefont {Y.}~\bibnamefont {Guo}}, \bibinfo {author}
  {\bibfnamefont {Y.}~\bibnamefont {Zhang}}, \bibinfo {author} {\bibfnamefont
  {Z.}~\bibnamefont {Yue}}, \bibinfo {author} {\bibfnamefont {J.-X.}\
  \bibnamefont {Yin}}, \bibinfo {author} {\bibfnamefont {M.}~\bibnamefont
  {Hashimoto}}, \bibinfo {author} {\bibfnamefont {D.}~\bibnamefont {Lu}},
  \bibinfo {author} {\bibfnamefont {S.}~\bibnamefont {Gorovikov}}, \bibinfo
  {author} {\bibfnamefont {P.}~\bibnamefont {Dai}}, \bibinfo {author}
  {\bibfnamefont {M.~Z.}\ \bibnamefont {Hasan}}, \bibinfo {author}
  {\bibfnamefont {Y.-P.}\ \bibnamefont {Feng}}, \bibinfo {author}
  {\bibfnamefont {R.~J.}\ \bibnamefont {Birgeneau}}, \bibinfo {author}
  {\bibfnamefont {Y.}~\bibnamefont {Shi}}, \bibinfo {author} {\bibfnamefont
  {C.-W.}\ \bibnamefont {Chu}}, \bibinfo {author} {\bibfnamefont
  {G.}~\bibnamefont {Chang}}, \bibinfo {author} {\bibfnamefont
  {Q.}~\bibnamefont {Si}}, \ and\ \bibinfo {author} {\bibfnamefont
  {M.}~\bibnamefont {Yi}},\ }\href@noop {} {\enquote {\bibinfo {title}
  {{Three-dimensional flat bands and Dirac cones in a pyrochlore
  superconductor, {T}o appear in {\it {N}at. {P}hys.}, {\em arXiv:2304.09066}
  (2023)}},}\ }\BibitemShut {NoStop}%
\bibitem [{\citenamefont {Yu}\ and\ \citenamefont {Si}(2012)}]{Yu_ss}%
  \BibitemOpen
  \bibfield  {author} {\bibinfo {author} {\bibfnamefont {R.}~\bibnamefont
  {Yu}}\ and\ \bibinfo {author} {\bibfnamefont {Q.}~\bibnamefont {Si}},\ }\href
  {\doibase 10.1103/PhysRevB.86.085104} {\bibfield  {journal} {\bibinfo
  {journal} {Phys. Rev. B}\ }\textbf {\bibinfo {volume} {86}},\ \bibinfo
  {pages} {085104} (\bibinfo {year} {2012})}\BibitemShut {NoStop}%
\bibitem [{\citenamefont {Yu}\ and\ \citenamefont {Si}(2017)}]{Yu-osmp17}%
  \BibitemOpen
  \bibfield  {author} {\bibinfo {author} {\bibfnamefont {R.}~\bibnamefont
  {Yu}}\ and\ \bibinfo {author} {\bibfnamefont {Q.}~\bibnamefont {Si}},\ }\href
  {\doibase 10.1103/PhysRevB.96.125110} {\bibfield  {journal} {\bibinfo
  {journal} {Phys. Rev. B}\ }\textbf {\bibinfo {volume} {96}},\ \bibinfo
  {pages} {125110} (\bibinfo {year} {2017})}\BibitemShut {NoStop}%
\bibitem [{\citenamefont {Komijani}\ and\ \citenamefont
  {Kotliar}(2017)}]{Komijani-osmp17}%
  \BibitemOpen
  \bibfield  {author} {\bibinfo {author} {\bibfnamefont {Y.}~\bibnamefont
  {Komijani}}\ and\ \bibinfo {author} {\bibfnamefont {G.}~\bibnamefont
  {Kotliar}},\ }\href {\doibase 10.1103/PhysRevB.96.125111} {\bibfield
  {journal} {\bibinfo  {journal} {Phys. Rev. B}\ }\textbf {\bibinfo {volume}
  {96}},\ \bibinfo {pages} {125111} (\bibinfo {year} {2017})}\BibitemShut
  {NoStop}%
\bibitem [{\citenamefont {Seki}\ and\ \citenamefont
  {Yunoki}(2017)}]{Yunoki2017}%
  \BibitemOpen
  \bibfield  {author} {\bibinfo {author} {\bibfnamefont {K.}~\bibnamefont
  {Seki}}\ and\ \bibinfo {author} {\bibfnamefont {S.}~\bibnamefont {Yunoki}},\
  }\href {\doibase 10.1103/PhysRevB.96.085124} {\bibfield  {journal} {\bibinfo
  {journal} {Phys. Rev. B}\ }\textbf {\bibinfo {volume} {96}},\ \bibinfo
  {pages} {085124} (\bibinfo {year} {2017})}\BibitemShut {NoStop}%
\bibitem [{\citenamefont {Setty}\ \emph
  {et~al.}(2023{\natexlab{c}})\citenamefont {Setty}, \citenamefont {Xie},
  \citenamefont {Sur}, \citenamefont {Chen}, \citenamefont {Vergniory},\ and\
  \citenamefont {Si}}]{Setty_v2}%
  \BibitemOpen
  \bibfield  {author} {\bibinfo {author} {\bibfnamefont {C.}~\bibnamefont
  {Setty}}, \bibinfo {author} {\bibfnamefont {F.}~\bibnamefont {Xie}}, \bibinfo
  {author} {\bibfnamefont {S.}~\bibnamefont {Sur}}, \bibinfo {author}
  {\bibfnamefont {L.}~\bibnamefont {Chen}}, \bibinfo {author} {\bibfnamefont
  {M.~G.}\ \bibnamefont {Vergniory}}, \ and\ \bibinfo {author} {\bibfnamefont
  {Q.}~\bibnamefont {Si}},\ }\href@noop {} {\bibfield  {journal} {\bibinfo
  {journal} {arXiv preprint arXiv:2309.14340}\ } (\bibinfo {year}
  {2023}{\natexlab{c}})}\BibitemShut {NoStop}%
\bibitem [{\citenamefont {Blason}\ and\ \citenamefont
  {Fabrizio}(2023)}]{Blason2023}%
  \BibitemOpen
  \bibfield  {author} {\bibinfo {author} {\bibfnamefont {A.}~\bibnamefont
  {Blason}}\ and\ \bibinfo {author} {\bibfnamefont {M.}~\bibnamefont
  {Fabrizio}},\ }\href {\doibase 10.1103/PhysRevB.108.125115} {\bibfield
  {journal} {\bibinfo  {journal} {Phys. Rev. B}\ }\textbf {\bibinfo {volume}
  {108}},\ \bibinfo {pages} {125115} (\bibinfo {year} {2023})}\BibitemShut
  {NoStop}%
\bibitem [{\citenamefont {Peralta~Gavensky}\ \emph {et~al.}(2023)\citenamefont
  {Peralta~Gavensky}, \citenamefont {Sachdev},\ and\ \citenamefont
  {Goldman}}]{Gavensky2023}%
  \BibitemOpen
  \bibfield  {author} {\bibinfo {author} {\bibfnamefont {L.}~\bibnamefont
  {Peralta~Gavensky}}, \bibinfo {author} {\bibfnamefont {S.}~\bibnamefont
  {Sachdev}}, \ and\ \bibinfo {author} {\bibfnamefont {N.}~\bibnamefont
  {Goldman}},\ }\href {\doibase 10.1103/PhysRevLett.131.236601} {\bibfield
  {journal} {\bibinfo  {journal} {Phys. Rev. Lett.}\ }\textbf {\bibinfo
  {volume} {131}},\ \bibinfo {pages} {236601} (\bibinfo {year}
  {2023})}\BibitemShut {NoStop}%
\bibitem [{\citenamefont {Zhao}\ \emph
  {et~al.}(2023{\natexlab{c}})\citenamefont {Zhao}, \citenamefont {Mai},
  \citenamefont {Bradlyn},\ and\ \citenamefont {Phillips}}]{Phillips2023}%
  \BibitemOpen
  \bibfield  {author} {\bibinfo {author} {\bibfnamefont {J.}~\bibnamefont
  {Zhao}}, \bibinfo {author} {\bibfnamefont {P.}~\bibnamefont {Mai}}, \bibinfo
  {author} {\bibfnamefont {B.}~\bibnamefont {Bradlyn}}, \ and\ \bibinfo
  {author} {\bibfnamefont {P.}~\bibnamefont {Phillips}},\ }\href {\doibase
  10.1103/PhysRevLett.131.106601} {\bibfield  {journal} {\bibinfo  {journal}
  {Phys. Rev. Lett.}\ }\textbf {\bibinfo {volume} {131}},\ \bibinfo {pages}
  {106601} (\bibinfo {year} {2023}{\natexlab{c}})}\BibitemShut {NoStop}%
\bibitem [{\citenamefont {Fabrizio}(2022)}]{Fabrizio2022}%
  \BibitemOpen
  \bibfield  {author} {\bibinfo {author} {\bibfnamefont {M.}~\bibnamefont
  {Fabrizio}},\ }\href@noop {} {\bibfield  {journal} {\bibinfo  {journal}
  {Nature Communications}\ }\textbf {\bibinfo {volume} {13}},\ \bibinfo {pages}
  {1} (\bibinfo {year} {2022})}\BibitemShut {NoStop}%
\bibitem [{\citenamefont {Wagner}\ \emph
  {et~al.}(2023{\natexlab{b}})\citenamefont {Wagner}, \citenamefont {Guerci},
  \citenamefont {Millis},\ and\ \citenamefont {Sangiovanni}}]{Sangiovanni_2}%
  \BibitemOpen
  \bibfield  {author} {\bibinfo {author} {\bibfnamefont {N.}~\bibnamefont
  {Wagner}}, \bibinfo {author} {\bibfnamefont {D.}~\bibnamefont {Guerci}},
  \bibinfo {author} {\bibfnamefont {A.~J.}\ \bibnamefont {Millis}}, \ and\
  \bibinfo {author} {\bibfnamefont {G.}~\bibnamefont {Sangiovanni}},\
  }\href@noop {} {\bibfield  {journal} {\bibinfo  {journal} {arXiv preprint
  arXiv:2312.13226}\ } (\bibinfo {year} {2023}{\natexlab{b}})}\BibitemShut
  {NoStop}%
\bibitem [{\citenamefont {Bollmann}\ \emph {et~al.}(2023)\citenamefont
  {Bollmann}, \citenamefont {Setty}, \citenamefont {Seifert},\ and\
  \citenamefont {K{\"o}nig}}]{Bollmann}%
  \BibitemOpen
  \bibfield  {author} {\bibinfo {author} {\bibfnamefont {S.}~\bibnamefont
  {Bollmann}}, \bibinfo {author} {\bibfnamefont {C.}~\bibnamefont {Setty}},
  \bibinfo {author} {\bibfnamefont {U.~F.}\ \bibnamefont {Seifert}}, \ and\
  \bibinfo {author} {\bibfnamefont {E.~J.}\ \bibnamefont {K{\"o}nig}},\
  }\href@noop {} {\bibfield  {journal} {\bibinfo  {journal} {arXiv preprint
  arXiv:2312.14926}\ } (\bibinfo {year} {2023})}\BibitemShut {NoStop}%
\bibitem [{\citenamefont {Yang}\ \emph {et~al.}(2015)\citenamefont {Yang},
  \citenamefont {Morimoto},\ and\ \citenamefont {Furusaki}}]{Yang2015}%
  \BibitemOpen
  \bibfield  {author} {\bibinfo {author} {\bibfnamefont {B.-J.}\ \bibnamefont
  {Yang}}, \bibinfo {author} {\bibfnamefont {T.}~\bibnamefont {Morimoto}}, \
  and\ \bibinfo {author} {\bibfnamefont {A.}~\bibnamefont {Furusaki}},\ }\href
  {\doibase 10.1103/PhysRevB.92.165120} {\bibfield  {journal} {\bibinfo
  {journal} {Phys. Rev. B}\ }\textbf {\bibinfo {volume} {92}},\ \bibinfo
  {pages} {165120} (\bibinfo {year} {2015})}\BibitemShut {NoStop}%
\bibitem [{\citenamefont {de'Medici}\ \emph {et~al.}(2005)\citenamefont
  {de'Medici}, \citenamefont {Georges},\ and\ \citenamefont
  {Biermann}}]{Medici_ss}%
  \BibitemOpen
  \bibfield  {author} {\bibinfo {author} {\bibfnamefont {L.}~\bibnamefont
  {de'Medici}}, \bibinfo {author} {\bibfnamefont {A.}~\bibnamefont {Georges}},
  \ and\ \bibinfo {author} {\bibfnamefont {S.}~\bibnamefont {Biermann}},\
  }\href {\doibase 10.1103/PhysRevB.72.205124} {\bibfield  {journal} {\bibinfo
  {journal} {Phys. Rev. B}\ }\textbf {\bibinfo {volume} {72}},\ \bibinfo
  {pages} {205124} (\bibinfo {year} {2005})}\BibitemShut {NoStop}%
\bibitem [{\citenamefont {Yu}\ \emph {et~al.}(2011)\citenamefont {Yu},
  \citenamefont {Zhu},\ and\ \citenamefont {Si}}]{Yu-cluster11}%
  \BibitemOpen
  \bibfield  {author} {\bibinfo {author} {\bibfnamefont {R.}~\bibnamefont
  {Yu}}, \bibinfo {author} {\bibfnamefont {J.-X.}\ \bibnamefont {Zhu}}, \ and\
  \bibinfo {author} {\bibfnamefont {Q.}~\bibnamefont {Si}},\ }\href {\doibase
  10.1103/PhysRevLett.106.186401} {\bibfield  {journal} {\bibinfo  {journal}
  {Phys. Rev. Lett.}\ }\textbf {\bibinfo {volume} {106}},\ \bibinfo {pages}
  {186401} (\bibinfo {year} {2011})}\BibitemShut {NoStop}%
\bibitem [{\citenamefont {Zhao}\ and\ \citenamefont
  {Paramekanti}(2007)}]{Zhao-cluster07}%
  \BibitemOpen
  \bibfield  {author} {\bibinfo {author} {\bibfnamefont {E.}~\bibnamefont
  {Zhao}}\ and\ \bibinfo {author} {\bibfnamefont {A.}~\bibnamefont
  {Paramekanti}},\ }\href {\doibase 10.1103/PhysRevB.76.195101} {\bibfield
  {journal} {\bibinfo  {journal} {Phys. Rev. B}\ }\textbf {\bibinfo {volume}
  {76}},\ \bibinfo {pages} {195101} (\bibinfo {year} {2007})}\BibitemShut
  {NoStop}%
\bibitem [{\citenamefont {Florens}\ and\ \citenamefont
  {Georges}(2004)}]{Florens2004}%
  \BibitemOpen
  \bibfield  {author} {\bibinfo {author} {\bibfnamefont {S.}~\bibnamefont
  {Florens}}\ and\ \bibinfo {author} {\bibfnamefont {A.}~\bibnamefont
  {Georges}},\ }\href {\doibase 10.1103/PhysRevB.70.035114} {\bibfield
  {journal} {\bibinfo  {journal} {Phys. Rev. B}\ }\textbf {\bibinfo {volume}
  {70}},\ \bibinfo {pages} {035114} (\bibinfo {year} {2004})}\BibitemShut
  {NoStop}%
\bibitem [{\citenamefont {Ding}\ \emph {et~al.}(2019)\citenamefont {Ding},
  \citenamefont {Yu}, \citenamefont {Si},\ and\ \citenamefont
  {Abrahams}}]{Ding-rotor2019}%
  \BibitemOpen
  \bibfield  {author} {\bibinfo {author} {\bibfnamefont {W.}~\bibnamefont
  {Ding}}, \bibinfo {author} {\bibfnamefont {R.}~\bibnamefont {Yu}}, \bibinfo
  {author} {\bibfnamefont {Q.}~\bibnamefont {Si}}, \ and\ \bibinfo {author}
  {\bibfnamefont {E.}~\bibnamefont {Abrahams}},\ }\href {\doibase
  10.1103/PhysRevB.100.235113} {\bibfield  {journal} {\bibinfo  {journal}
  {Phys. Rev. B}\ }\textbf {\bibinfo {volume} {100}},\ \bibinfo {pages}
  {235113} (\bibinfo {year} {2019})}\BibitemShut {NoStop}%
\end{thebibliography}%

\clearpage
\onecolumngrid
\appendix

\section{SUPPLEMENTARY MATERIAL}

\section{Noninteracting Hamiltonian and Dirac nodes}
\label{appsec:model}
\begin{figure}[h]
    \centering
    \includegraphics[width=0.4\linewidth]{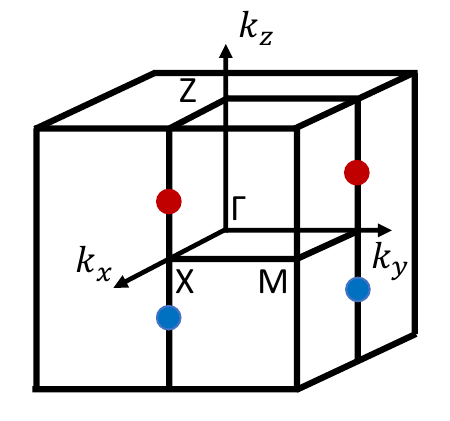}
    \caption{ The position of the Dirac points in the noninteracting limit. The red (blue) indicates a positive (negative) sign of monopole charge (defined for spin-$\up$ component). 
    }
    \label{fig:nodes}
\end{figure}

In this section, we 
present the further details on the kinetic part of the Hamitlonian.
As described in the main text, 
it preserves the $U(1)$-spin rotational symmetry and, therefore $\mathcal{H}_0^{\up} = [\mathcal{H}_0^{\up}]^{\dagger}$. In the basis of $\Psi_{\kbf}^{\up}=[\psi_{A h}^{\up},\psi_{B h}^{\up}, \psi_{A l}^{\up},\psi_{B l}^{\up}]$, the Hamiltonian is written as 
\begin{equation}
    H_0^{\up}(\kbf) = \mx
    t_{h} \epsilon_{\kbf} & t^{soc}_h V_{\kbf} & t_{hl} \epsilon_{\kbf} & 0 \\
    t^{soc}_h V_{\kbf}^* &  -t_{h} \epsilon_{\kbf} & 0 & t_{hl} \epsilon_{\kbf}\\
    t_{hl} \epsilon_{\kbf} & 0 & t_{l} \epsilon_{\kbf} & t^{soc}_l V_{\kbf}  \\
    0 & t_{hl} \epsilon_{\kbf} & t_{l}^{soc}V_{\kbf}^{*} & -t_l \epsilon_{\kbf}
    \ex .
\end{equation}
The parameters $t_{h/l}$, $t_{h/l}^{soc}$ and $t_{hl}$ are chosen to be real, and 
\begin{align}
    \epsilon_{\kbf} &= 2(\cos k_x + \cos k_y + \cos k_z) \, , \\
    V_{\kbf} & = 2(\sin k_x - i\sin k_y) 
\end{align}
are the form factors related to the nearest neighbor direct hopping and spin-orbit coupling, respectively.
Altogether, there are $8$ Dirac nodes,
with $4$ each coming from the heavy (light) orbitals. They are sitting at the momenta $\kbf_*=(\pi,0,\pm \pi/2)$ and $\kbf_*=(0,\pi,\pm \pi/2)$, with the sign of the monopole charge (defined for spin-$\up$) as shown in Fig.~\ref{fig:nodes}. 
The model we 
consider 
has a $U(1)$ spin rotational symmetry (along with time-reversal and inversion symmetries);
as a result, the monopole charge of a Dirac point can be defined from a single spin species.
Dirac points with other kinds of symmetries can also be considered~\cite{Yang2015}.

\section{Details of the cluster slave spin calculation}\label{appsec:ss}

In this section, we describe the cluster SS method.
As we consider only the interactions on the heavy orbitals, the physical fermionic operators of the heavy orbitals are 
expressed in terms of the SS and spinon operators as $\psi_{\tau h\sigma}^{\dagger} = S^+_{\tau h\sigma}f_{\tau h\sigma}^{\dagger}$. The intra-orbital Hubbard interaction can be represented by the SS operators as:
\begin{equation}
    \mathcal{H}_{int} =  \frac{U}{2} \sum_{i,\tau} \left(\sum_{\sigma} S^z_{\tau h\sigma}\right)^2 \, .
\end{equation}
This  representation has
a U(1) symmetry.
Compared to the $Z_2$ formulation
\cite{Medici_ss}, 
it has the advantage of 
allowing the charge degrees of freedom to be expressed in terms of the SS operators alone.
Finally, compared to the cluster formulation~\cite{Yu-cluster11,Zhao-cluster07} of the 
slave rotor approach~\cite{Florens2004,Ding-rotor2019}, the small Hilbert space of a slave spin operator facilitates the analysis of the cluster
equations.

\begin{figure}[h]
    \centering
    \includegraphics[width=0.6\linewidth]{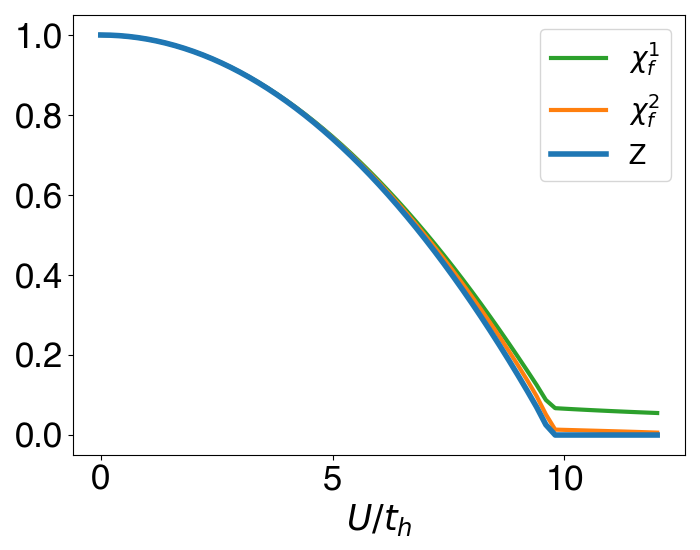}
    \caption{The quasiparticle weight and the expectation values of the bond operators $\chi_f^1$ and $\chi_f^2$ versus the interaction strength.
    }
    \label{fig:Qf}
\end{figure}

We then treat the full Hamiltonian
at the saddle point level,
which 
contains two decomposed 
effective Hamiltonian with the SS and spinon operactors only. They are specified as follows,
\begin{equation}
\begin{aligned}
    \mathcal{H}^f &= \sum_{\substack{ij, \sigma, \tau\tau'} }t_{\tau\tau'}^he^{i\theta_{ij,h\sigma}^{\tau\tau'}} \langle S^+_{i\tau h \sigma} S^-_{j\tau'h\sigma} \rangle f_{i\tau h\sigma}^{\dagger}f_{j\tau' h\sigma} + h.c.  + \sum_{\substack{ij,\sigma,\tau\tau'}} t_{hl} \langle S_{i\tau h\sigma}^+\rangle f_{i\tau h\sigma}^{\dagger}\psi_{j\tau'l\sigma} + h.c.\\
    & - \lambda \sum_{i\sigma,\tau }f_{i\tau h\sigma}^{\dagger}f_{i\tau h\sigma}
    +\mathcal{H}[\psi_{Al}, \psi_{Bl} ]
    \, ,
\end{aligned}
\end{equation}
\begin{equation}
\begin{aligned}
    \mathcal{H}^S &= \sum_{\substack{ij\sigma,
    \tau\tau' }} t_{\tau\tau'}^h\langle e^{i\theta_{ij,h\sigma}^{\tau\tau'}}  f_{i\tau h\sigma}^{\dagger} f_{j\tau' h\sigma} \rangle S^+_{i\tau h\sigma} S^-_{j\tau' h\sigma} + h.c. + \sum_{\substack{ij\sigma, \tau \tau' } } t_{hl} \langle f_{i\tau h\sigma}^{\dagger} \psi_{j\tau'l\sigma} \rangle S_{i\tau h\sigma}^+ + h.c.  \\
    & + \lambda \sum_{i\sigma, \tau } \left( S^z_{i\tau h\sigma} + \frac{1}{2} \right) + \mathcal{H}_{int}
    \, ,
\end{aligned}
\end{equation}
where $\theta^{\tau\tau'}_{ij,h\sigma}$ is the phase factor between the sublattices
$\tau$ and $\tau'$ of the heavy orbitals in unit cells $i$ and $j$ for spin $\sigma$, and $t_{A A}^h = - t_{B B}^h=t_h$, $t_{AB}^h=t_{BA}^h=t^{soc}_{h}$. $\mathcal{H}[\psi_{Al},\psi_{Bl} ]$ is the noninteracting Hamiltonian between the light orbitals. 
A Lagrangian multiplier  $\lambda$ is introduced to fix the constraint $S^z_{i\tau h\sigma}+1/2=f^{\dagger}_{i\tau h\sigma}f_{i\tau h\sigma}$.
Similar to the cluster formulations of other auxiliary-particle approaches \cite{Yu-cluster11,Zhao-cluster07},
we treat
the SS Hamiltonian by considering a cluster with two unit cells
embedded in an average medium. This
leads to a cluster SS Hamiltonian
\begin{equation}\label{eq:hs_cluster}
    \begin{aligned}
        \mathcal{H}^S&=\sum_{\tau} t_{\tau\tau}^h \chi^{\tau\tau}_S S^+_{\tau h\sigma,1}S^-_{\tau h\sigma,2} + h.c.+ \sum_{ \substack{\tau\tau', \tau\neq\tau'} } t_{\tau\tau'}^h \chi^{\tau\tau'}_S S^+_{\tau h\sigma,1} S^-_{\tau' h\sigma, 2} +h.c.\\
        &+ \sum_{\tau } 5t_{\tau\tau}^h \chi_{S}^{\tau\tau}\langle S_{\tau h\sigma}^+\rangle \left( S^-_{\tau h\sigma,1} + S^-_{\tau h\sigma,2} \right) +h.c. + \sum_{\substack{\tau\tau', \tau\neq\tau'} } 3t_{\tau\tau'}^h \chi^{\tau\tau'}_S \langle S_{\tau h\sigma}^+ \rangle \left( S^-_{\tau' h\sigma,1} + S^-_{\tau' h\sigma,2} \right) + h.c.\\
        & + \sum_{\substack{\sigma, \tau\tau' } } 6t_{hl} \chi^{f_{\tau}\psi_{\tau'} }_S \left( S^+_{\tau h\sigma,1} + S^+_{\tau h\sigma,2} \right)+ h.c. +\lambda \sum_{\sigma,\tau } \left( S^z_{\tau h\sigma, 1} + S^z_{\tau h\sigma, 2} + 1\right) + \mathcal{H}_{int} \, ,
    \end{aligned}
\end{equation}
where the subscripts $1,2$ label the two unit cells in the cluster, and
\begin{equation}
    \begin{aligned}
        \chi^{\tau\tau}_S &= \frac{1}{12N_{\kbf} } \sum_{\kbf} \epsilon_{\kbf} \langle f_{\kbf\tau h\sigma}^{\dagger}f_{\kbf \tau h\sigma} \rangle \, , \\
        \chi^{\tau\tau'}_S &= \frac{1}{8N_k} \sum_{\kbf} V_{\kbf} \langle f^{\dagger}_{\kbf\tau h\sigma} f_{\kbf\tau'h\sigma} \rangle \, , \\
        \chi^{f_{\tau}\psi_{\tau'} }_{S} &= \frac{1}{12N_k} \sum_{\kbf} \epsilon_k \langle f^{\dagger}_{\kbf\tau h\sigma} \psi_{\kbf\tau' l\sigma} \rangle \, . 
     \end{aligned}
\end{equation}
The expectation 
values appearing in the $\mathcal{H}^f$ are then replaced by the expectation values obtained from $\mathcal{H}^S$ of Eq.~(\ref{eq:hs_cluster}):
\begin{equation}\label{eq:op_17}
    \langle S^+_{i\tau h\sigma} S^-_{j\tau'h\sigma}\rangle = \begin{cases}
    \langle S_{\tau h\sigma,1}^+ S^-_{\tau'h\sigma,2}\rangle & i,j 
    \in n.n.\\
    0 & otherwise
    \end{cases}
\end{equation}
\begin{equation}\label{eq:op_18}
    \langle S^+_{i\tau h\sigma}\rangle = \langle S^+_{\tau h\sigma, 1}\rangle = \langle S^+_{\tau h\sigma, 2}\rangle = \sqrt{Z_{\tau h\sigma}} \, .
\end{equation}
The spinon Hamiltonian is 
then written as
\begin{equation}\label{eq:hf_19}
\begin{aligned}
    \mathcal{H}^f &= \sum_{ij\sigma, \tau} t_{\tau\tau}^h \chi^{1}_f f^{\dagger}_{i\tau h\sigma} f_{j\tau h\sigma} + h.c. + \sum_{\substack{ij\sigma, \tau\tau', \tau\neq \tau' }} t_{\tau\tau'}^h  \chi^{2}_f f_{i\tau h\sigma}^{\dagger}f_{j\tau h\sigma} + h.c. \\
    & + \sum_{\substack{ij\sigma, \tau
    \tau' } } t_{hl} \sqrt{Z_{i\tau h\sigma}} f^{\dagger}_{i\tau h\sigma}\psi_{j\tau' l\sigma} + h.c.- \lambda \sum_{i\sigma,\tau}f_{i\tau h\sigma}^{\dagger}f_{i\tau h\sigma}
    +\mathcal{H}[\psi_{Al}, \psi_{Bl} ],
\end{aligned}
\end{equation}
with $\chi_f^1=\langle S^+_{A h\sigma,1}S^-_{A h\sigma,2} \rangle$ and $\chi_f^2= \langle S^+_{A h\sigma,1}S^-_{B h\sigma,2} \rangle$. Combining the equations from Eq.~(\ref{eq:hs_cluster})-(\ref{eq:hf_19}), it is sufficient to solve the interacting Hamiltonian self-consistently. The variance of $\chi^1_f$, $\chi^2_f$ and quasipartice weight is shown in Fig.~\ref{fig:Qf}. Both $\chi^1_f$ and $\chi^2_f$ remain to be nonzero throughout, while $Z=0$ in the OSMP. We note that the filling of each orbital remains $1$ throughout the phase diagram.
 
\section{Single-particle Green's function 
in the cluster slave spin method}
\label{appsec:gf}

The cluster SS method gives rise to a translationally-invariant 
single-particle Green's function of physical electrons. 
It is obtained by
\begin{equation}
\begin{aligned}\label{eq:g_ss_2}
     G_{\psi}^{\tau\tau'} (\kbf, \omega) &= \sum_{\Rbf} e^{-i\kbf\Rbf} G_{\psi}^{\tau\tau'} (\Rbf, \omega) \\
     & = G_{\psi}^{\tau\tau'}(\bm{0}, \omega) + \sum_{\Rbf \neq \bm{0}
     } e^{-i\kbf\Rbf}G_{\psi}^{\tau\tau'}(\Rbf,\omega),
\end{aligned}
\end{equation}
where the spin and orbital indices are abbreviated.
At zero temperature, the first term is calculated by
\begin{equation}
    G_{\psi}^{\tau\tau'}(\bm{0},\omega) =\sum_{\kbf} \sum_{\alpha}\sum_m \Lambda_{\tau\tau'}^{\alpha,\kbf} \left[ \frac{\langle 0|S^+_{\tau,1} |m \rangle\langle m|S^-_{\tau',1}|0\rangle \left( 1- \Theta(\epsilon_{\kbf}^{\alpha}) \right) }{\omega-E_m-\epsilon^{\alpha}_{\kbf} } 
    +  \frac{\langle 0|S^-_{\tau',1} |m \rangle\langle m|S^+_{\tau,1}|0\rangle \Theta(\epsilon_{\kbf}^{\alpha}) }{\omega+E_m-\epsilon^{\alpha}_{\kbf} } \right],
\end{equation}
where $\Lambda_{\tau\tau'}^{\alpha,\kbf} = \langle\alpha_{\kbf}|f^{\dagger}_{\kbf\tau}f_{\kbf\tau'} |\alpha_{\kbf} \rangle $, with $\epsilon_{\kbf}^{\alpha} $ ($|\alpha_{\kbf} \rangle $) the $\alpha$-th eigenvalues (eigenvectors) of the spinon Hamiltonian. 
Here, $|m\rangle$ enumerates the eigenvectors of the cluster SS Hamiltonian (Eq.~(\ref{eq:hs_cluster})) and $E_m$ denotes the corresponding eigen-energy. 
The second term can be obtained from:
\begin{equation}\label{eq:g_ss_incoh}
    G_{\psi}^{\tau\tau'}(\Rbf,\omega) =\sum_{\kbf} \sum_{\alpha}\sum_m e^{i\kbf\Rbf}\Lambda_{\tau\tau'}^{\alpha,\kbf} \left[ \frac{\langle 0|S^+_{\tau,1} |m \rangle\langle m|S^-_{\tau',2}|0\rangle \left( 1- \Theta(\epsilon_{\kbf}^{\alpha}) \right) }{\omega-E_m-\epsilon^{\alpha}_{\kbf} } 
    +  \frac{\langle 0|S^-_{\tau',2} |m \rangle\langle m|S^+_{\tau,1}|0\rangle \Theta(\epsilon_{\kbf}^{\alpha}) }{\omega+E_m-\epsilon^{\alpha}_{\kbf} } \right],
\end{equation}
where the operator $S^{\pm}_{\tau,1}$ and $S^{\pm}_{\tau,2}$ are sitting on different unit cells in the SS cluster. 
The translational invariance of the Green's function is satisfied because all the sites and bonds pick up the SS contributions according to Eqs.~(\ref{eq:op_17}) and (\ref{eq:op_18}).

These two terms can be further separated into coherent and incoherent parts. The coherent part, associated directly with the quasiparticles weights as described by the first term in Eq.~(\ref{eq:g_ss}), is obtained from Eq.~(\ref{eq:g_ss_2})-(\ref{eq:g_ss_incoh}) with $m=0$. The incoherent part, on the other hand, is obtained from Eq.~(\ref{eq:g_ss_2})-(\ref{eq:g_ss_incoh}) by summing up terms with $m\neq0$.

\end{document}